\documentclass[final,5p,times,twocolumn]{elsarticle}
\usepackage{graphicx}
\usepackage{amssymb}
\usepackage{reviewphysics}
\journal{arxiv.org}

\begin{document}

\begin{frontmatter}

\title{Signal Characteristics of a Resistive-Strip Micromegas Detector with an Integrated Two-Dimensional Readout}


\author[A]{Tai-Hua Lin}
\author[A]{Andreas D\"udder}
\author[A]{Matthias Schott}
\cortext[mycorrespondingauthor]{Matthias Schott}
\ead{mschott@cern.ch}
\author[A]{Chrysostomos Valderanis}
\author[A]{Laura Wehner}
\author[A]{Robert Westenberger}
\address[A]{Institute of Physics, Johannes Gutenberg University Mainz, GERMANY}

\begin{abstract}

In recent years, micropattern gaseous detectors, which comprise a two-dimensional readout structure within one PCB layer, received significant attention in the development of precision and cost-effective tracking detectors in medium and high energy physics experiments. In this article, we present for the first time a systematic performance study of the signal characteristics of a resistive strip micromegas detector with a two-dimensional readout, based on test-beam and X-ray measurements. In particular, comparisons of the response of the two independent readout-layers regarding their signal shapes and signal reconstruction efficiencies are presented.

\end{abstract}


\end{frontmatter}


\section{\label{Sec:Intro} Introduction}

Since its invention in 1996 \cite{Giomataris:1995fq}, the technology of Micromegas detectors has been under constant development~\cite{Giomataris:2004aa}. In recent years, Micropattern gaseous detectors received significant attention in the development of precision and cost-effective large-scale tracking detectors fo high-energy physics experiments \cite{Alexopoulos:2009zza}, \cite{Wotschack:2013ola}. The first Micromegas detectors with a two-dimensional readout structure integrated within one PCB board and a resistive strip layer, have been developed within the RD51 \cite{Chefdeville:2011zz} and MAMMA (Muon ATLAS Micromegas Activity) collaborations \cite{Alexopoulos:2010zz}. Micromegas detectors with such a 2D readout implementation have the advantage of a reduced material budget, and even more importantly for a fixed mechanical structure of two readout-layers. 

\begin{figure*}
\begin{minipage}{0.49\textwidth}
\resizebox{1.0\textwidth}{!}{\includegraphics{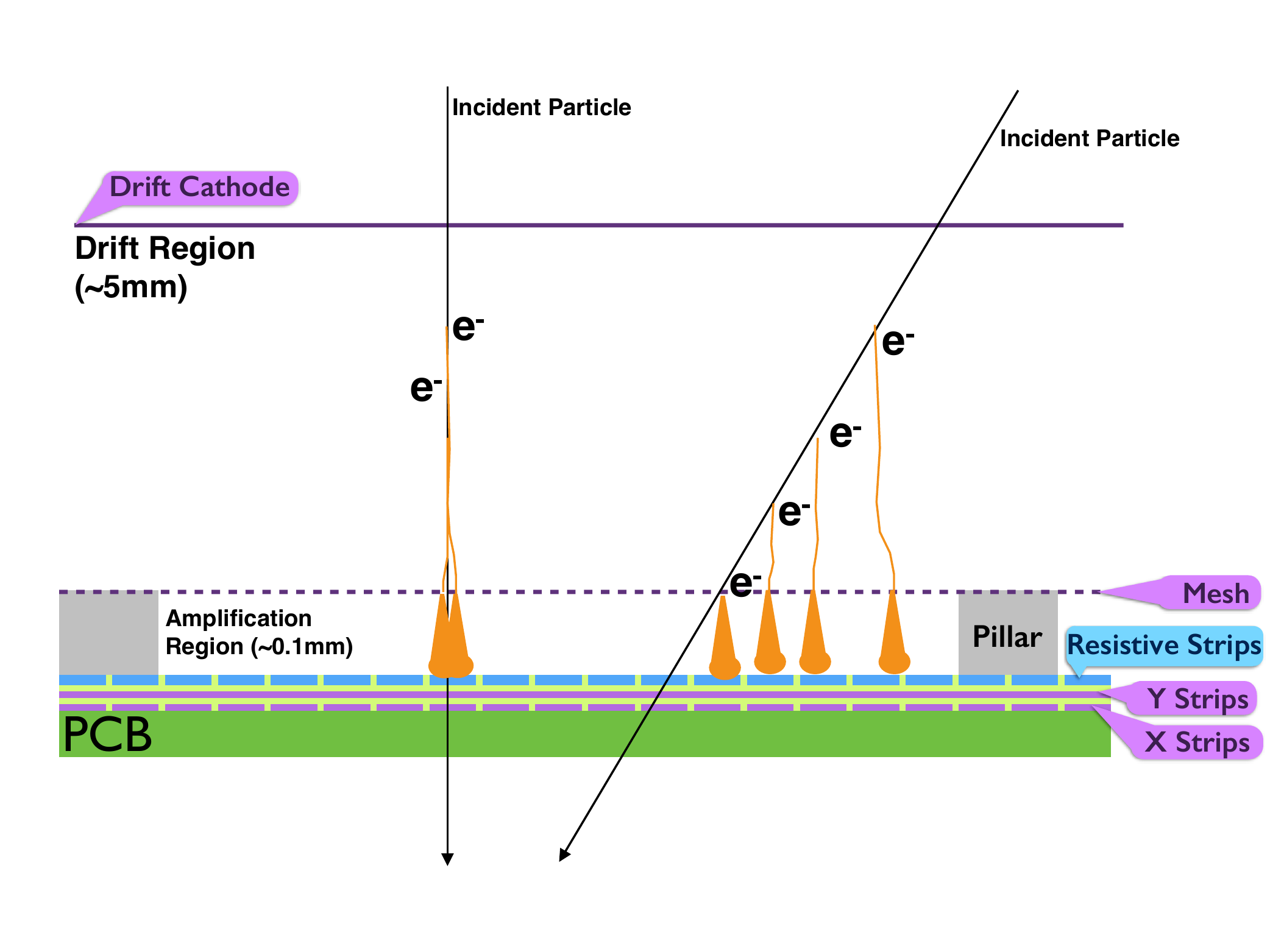}}
\caption{\label{fig:PincipleMicro}Illustration of basic principle of a Micromegas detector, with two incident ionizing particles. The incident particles ionize gas atoms, and the resulting secondary electrons drift to the lower mesh where they initiate to an electron cascade that can be measured on the readout strips.}
\end{minipage}
\hspace{0.5cm}
\begin{minipage}{0.49\textwidth}
\resizebox{1.0\textwidth}{!}{\includegraphics{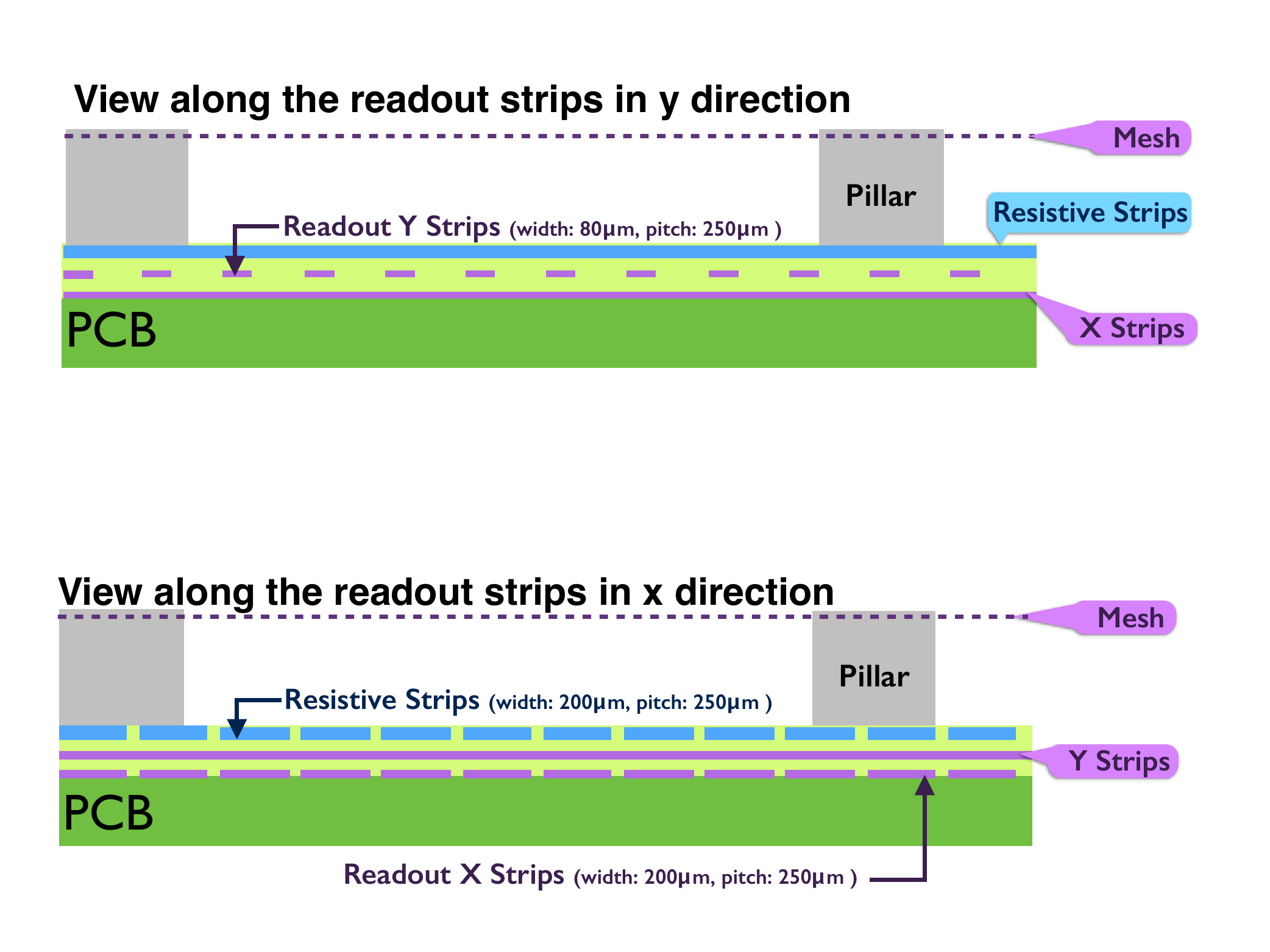}}
\caption{\label{Fig:DetLayout} Illustration of a spark protected Micromegas detector with two-dimensional layout. View along the readout strips in y- (upper) and x-direction (lower). The resistive strips cover only the upper readout-layer, while the second readout layer is separated with isolating material.}
\end{minipage}
\end{figure*}

A typical layout of a Micromegas detector is shown in Figure \ref{fig:PincipleMicro}. A planar drift electrode and a readout electrode are separated by a gap of a few $\mathrm{mm}$. While the drift electrode is covered uniformly with a conducting layer, e.g. copper, the readout electrode is usually made of a PCB board with a uniform structure of conductors, separated by insulating material. The strip width and the distance between strips can be chosen depending on the final application. The gap between the two electrodes is filled with an ionization gas, e.g. a 90:10 mixture of Argon and $\mathrm{CO_2}$. A metal mesh is placed at 50-100$\,\mathrm{\mu m}$ above the readout electrode, defining two volumes. The volume between the drift electrode and the mesh is called drift volume, the volume between mesh and PCB is the amplification volume. In order to create the drift and amplification environment, two external voltages are applied to the drift cathode and the resistive strips. 
A typical electric field in the drift volume is $\approx 600\,\mathrm{V/cm}$. A significantly higher electric field of $\approx 50\,\mathrm{kV/cm}$ is reached in the amplification volume. This high electric field might lead to sparks, resulting in dead-time and potential damage to the detector and the subsequent readout-system. Therefore, the readout strips are covered with a resistive protection layer \cite{Alexopoulos:2011zz} that consists of a thin insulator with a resistive paste. The resistive protection layer has a resistivity in the order of $\sim \mathrm{M\Omega/cm}$, and usually matches the geometry of the readout electrodes, in order to minimize a charge spread over several readout strips. 

A charged particle that enters the drift volume ionizes gas atoms, generating free electrons. These electrons drift along the electric field lines to the mesh with a typical velocity of $\approx 5\,\mathrm{cm/\mu s}$. The mesh appears transparent to the drift electrons because of the larger electric field in the subsequent amplification volume. Once the drift electrons reach the high electric field in the amplification volume, they are accelerated sufficiently to cause a cascade of secondary electrons (avalanche) leading to an amplification factor of $\approx 10^4$ within $1\,\mathrm{ns}$. These secondary electrons reach the resistive layer and induce a signal on the readout electrodes, via a capacitive coupling to resistive strips. A detailed introduction to the signal formation in Micromegas detectors can be found in \cite{Dris:2014qpa}.

In this paper, we describe the performance of a Micromegas detector with a two-dimensional readout structure.  A schematic layout is illustrated in Figure \ref{Fig:DetLayout}. In contrast to conventional Micromegas detectors, this layout foresees two independent readout electrodes in orthogonal directions (denoted as x and y in the following), printed onto the same PCB. While previous studies \cite{Byszewski:2012zz} focused on the performance of those detector layouts for one specific configuration, we present here a comprehensive list of signal characteristics on both layers results in dependence of the chosen gas-mixture, amplification-voltages ($V_A$) and drift-voltages ($V_D$). In particular, we compare the signal identification efficiency of both layers with respect to each other and the differences in the corresponding signal shapes.

The paper is structured as follows: In section \ref{Sec:Detector}, the detector layout and the experimental setups at the MAMI (Mainzer Mikrotron) accelerator and for X-ray measurements 
are described. The signal characteristics are described in section \ref{Sec:Signal}, while the efficiency studies are presented in section \ref{Sec:Efficiency}. A summary and concluding remarks can be found in section \ref{Sec:Summary}.

\section{\label{Sec:Detector}Prototype Micromegas Detector and Experimental Conditions}

The test chamber used for the studies presented here is based on a previous detector design by the MAMMA collaboration \cite{Byszewski:2012zz}, which focuses on the development and test of large-area muon detectors for the upgrade program of the ATLAS experiment. The readout electrode comprises 360 copper readout strips in each of the x- and the y-direction, where the strip pitch is $250\,\mathrm{\mu m}$ for both layers. The readout strips of the upper layer (defined as y-layer) are printed directly on top of the PCB and are covered by the resistive strips with a resistivity of $\approx 20\,\mathrm{M\Omega/cm}$. The lower layer (defined as x-layer) is separated from the upper layer by $70\,\mathrm{\mu m}$ of $FR4$, i.e. the same material used as isolating material in the PCB. The readout strips of the x-layer have a width of $200\,\mathrm{\mu m}$ and are placed parallel to the resistive strips, while the strips in the y-layer have a width of $80\,\mathrm{\mu m}$ and are perpendicular placed to the resistive strips. The larger width of the x-layer readout strips compensates for the weaker capacative coupling to the strips of the x-layer. 

The amplification mesh is made of woven stainless steel of 157 lines/cm with a line-diameter of $18\,\mathrm{\mu m}$. It is mounted on support pillars of $0.4\,\mathrm{mm}$ diameter and covers an area of $10\times10\,\mathrm{cm^2}$ which defines the active area of the detector. The support pillars are placed along a regular matrix with $2.5\,$mm spacing in both directions. 

\begin{figure*}[t]
\resizebox{0.47\textwidth}{!}{\includegraphics{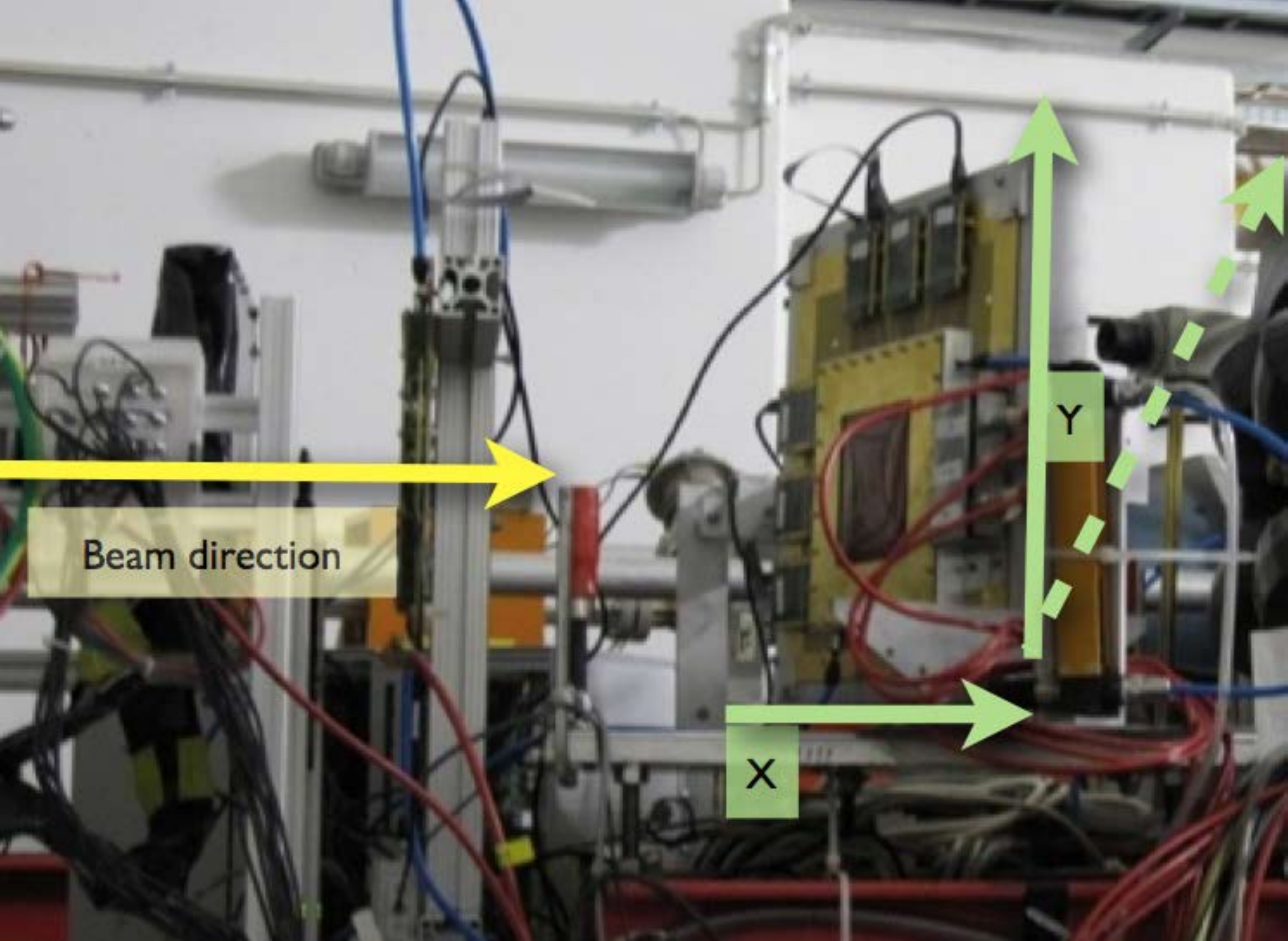}}
\hspace{0.5cm}
\resizebox{0.46\textwidth}{!}{\includegraphics{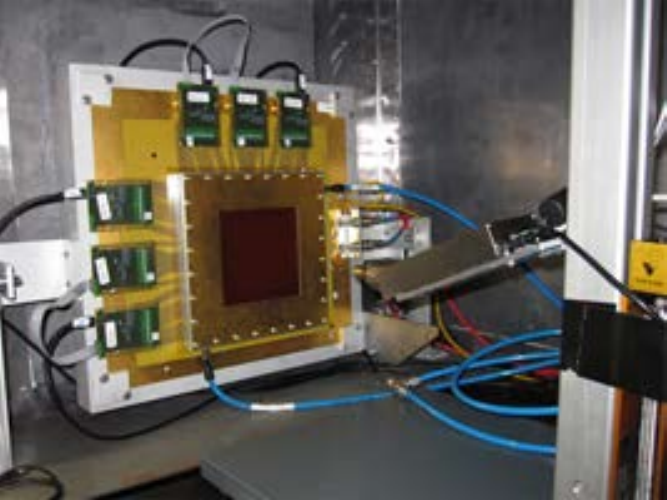}}
\caption{\label{Fig:DetImage}Photo of the test-beam setup at the MAMI accelerator (left) and the X-ray measurement setup with the prototype chamber (right.)}
\resizebox{0.495\textwidth}{!}{\includegraphics{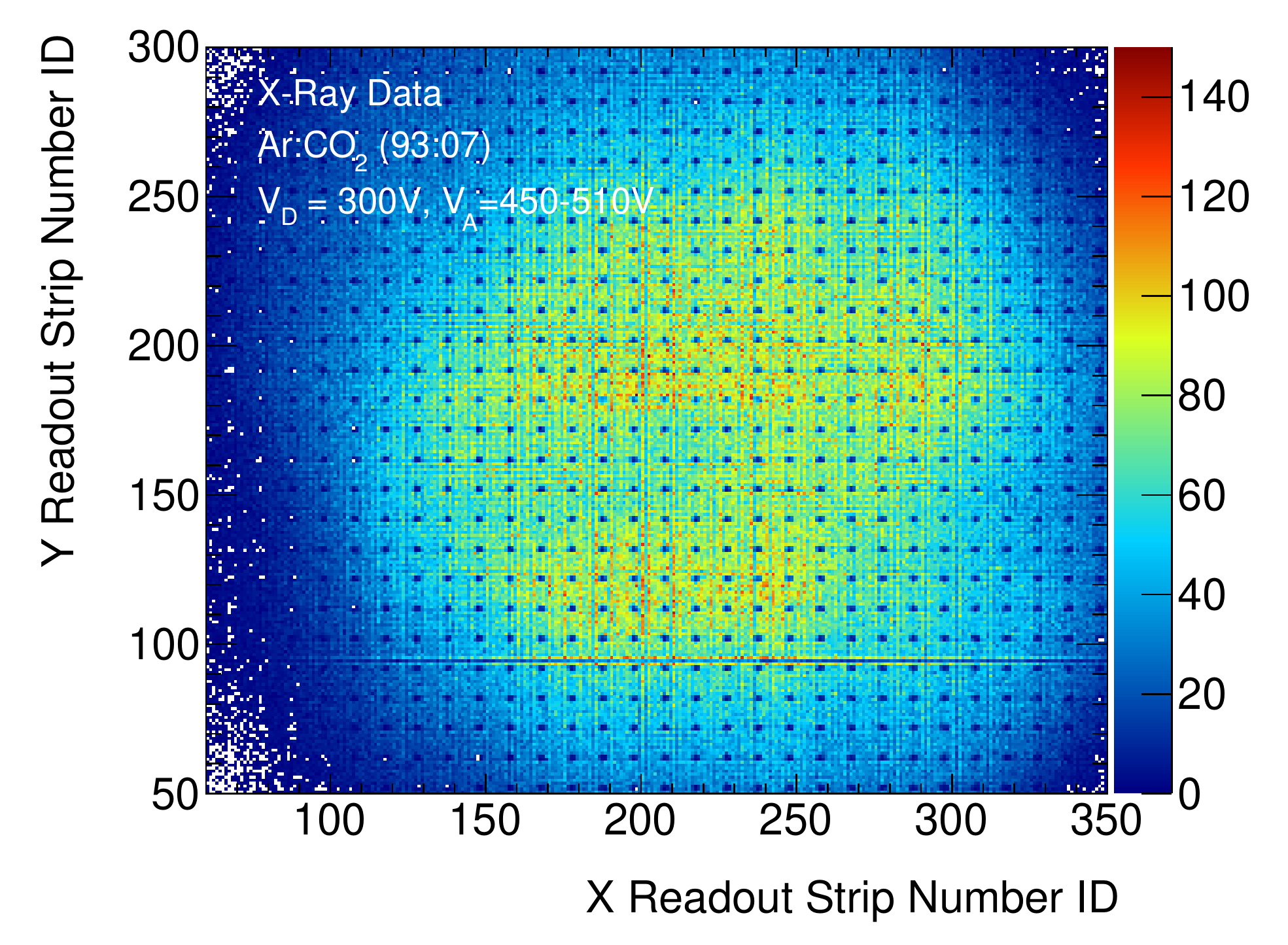}}
\resizebox{0.495\textwidth}{!}{\includegraphics{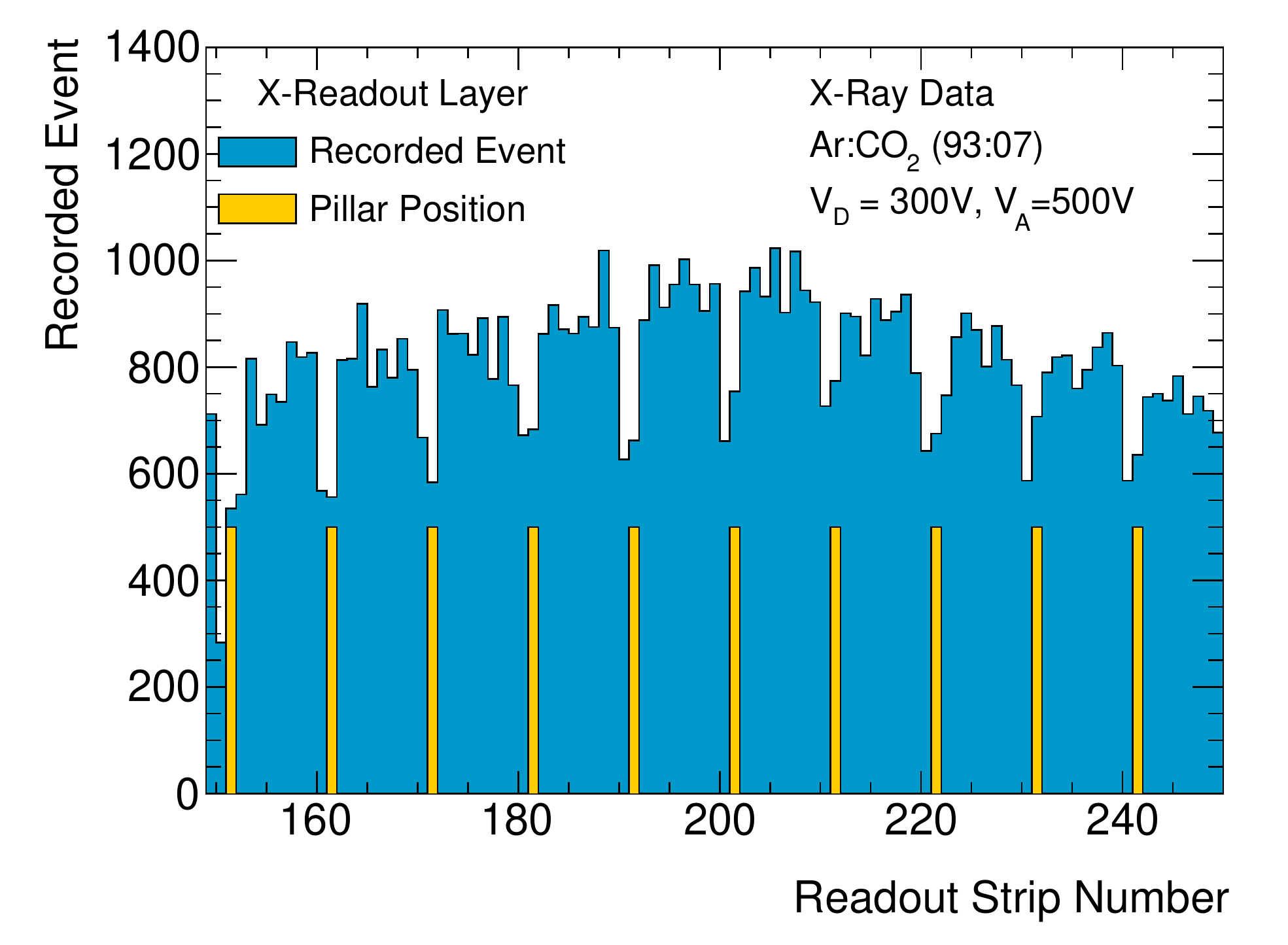}}
\caption{\label{fig:EventDisplayPillar}Measured distribution of reconstructed events during the X-ray runs in both layers (left) and for the X-readout layer (right). The readout strip with the maximal recorded charge has been choosen for each event. The support pillar structure is clearly visible.}
\end{figure*}

The data acquisition is based on the RD51 Scalable Readout System (SRS) \cite{Martoiu:2013aca, Martoiu:2011zja}. The signalprocessing on the detector is based on the Analog Pipeline Voltage chip with $0.25\,\mathrm{\mu m}$ CMOS technology (APV25) \cite{APVJones} where the analog signal data is transmitted via HDMI cables to SRS electronics. The SRS electronics process the analog signal which is then further analyzed. It should be noted that the heigth of the recorded readout signal can be interpreted as a measure of the charge on a given readout strip at a given time.

The test-beam measurements of the chamber have been conducted in August 2013 at the MAMI accelorator facility at the Johannes Gutenberg University Mainz. MAMI provides a quasi continues electron beam with energy up to $1.5\,\GeV$. A beam energy of $850\,\MeV$ was used for the measurement in this paper. Because of the operation conditions available at the test-beam, an Ar:$\mathrm{CO_2}$ gas mixture of 70:30 had to be used for the measurements. This is not an optimal value for the operation of micromegas chambers; however, clean signals have been recorded by applying drift and amplification voltages of $V_D=230\,$V and $V_A=680\,$V, respectively. The detector was placed in nine different orientations to the test-beam (Figure \ref{Fig:DetImage}) in order to study the performance for different incident angles of primary particles. The beam axis was chosen to be perpendicular to the readout-panel for reference measurements. Then the orientation was changed by $20^\circ$ and $40^\circ$, independently for the x and y axes.


\begin{figure*}[th]
\resizebox{0.495\textwidth}{!}{\includegraphics{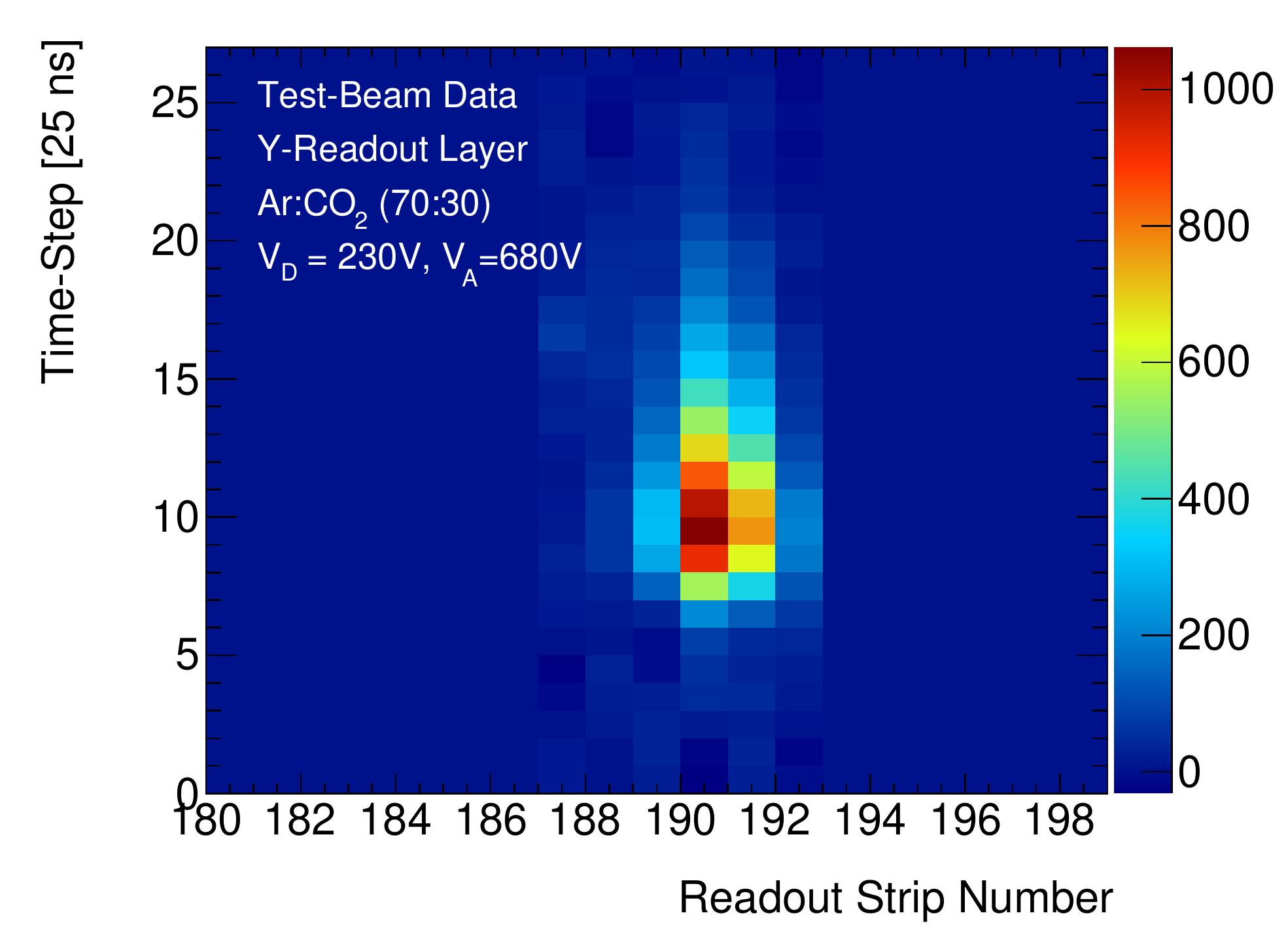}}
\resizebox{0.495\textwidth}{!}{\includegraphics{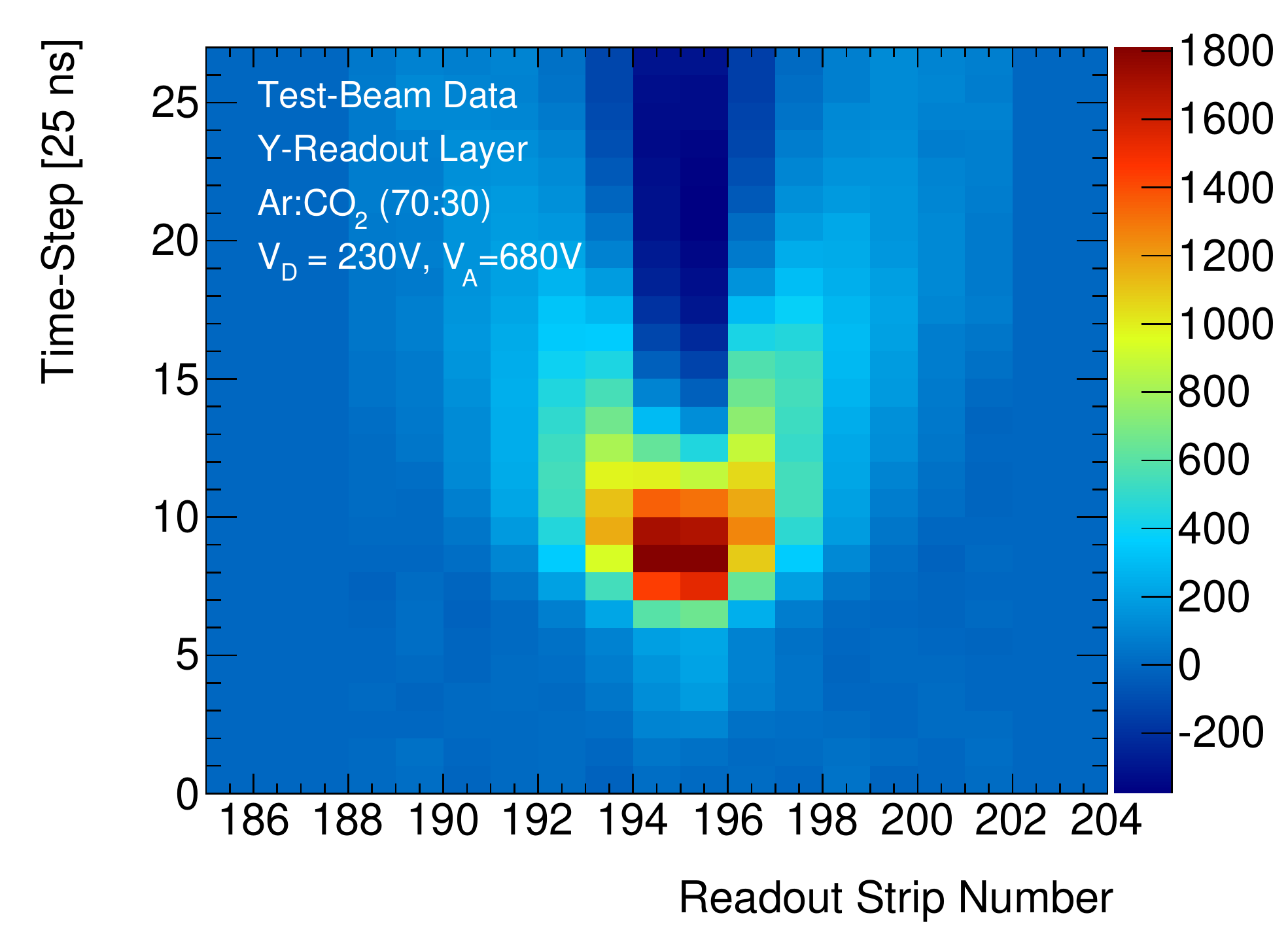}}
\resizebox{0.495\textwidth}{!}{\includegraphics{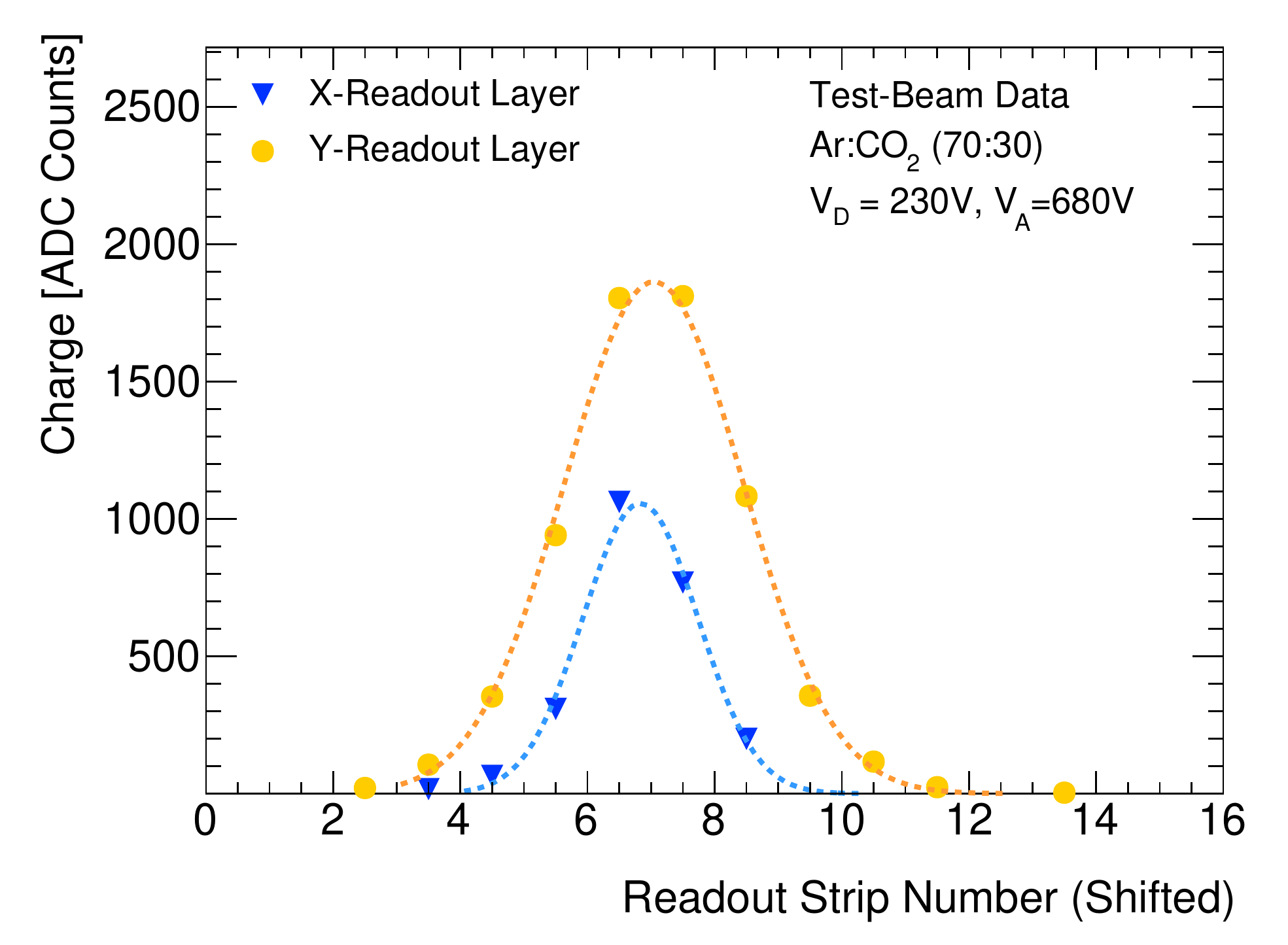}}
\resizebox{0.495\textwidth}{!}{\includegraphics{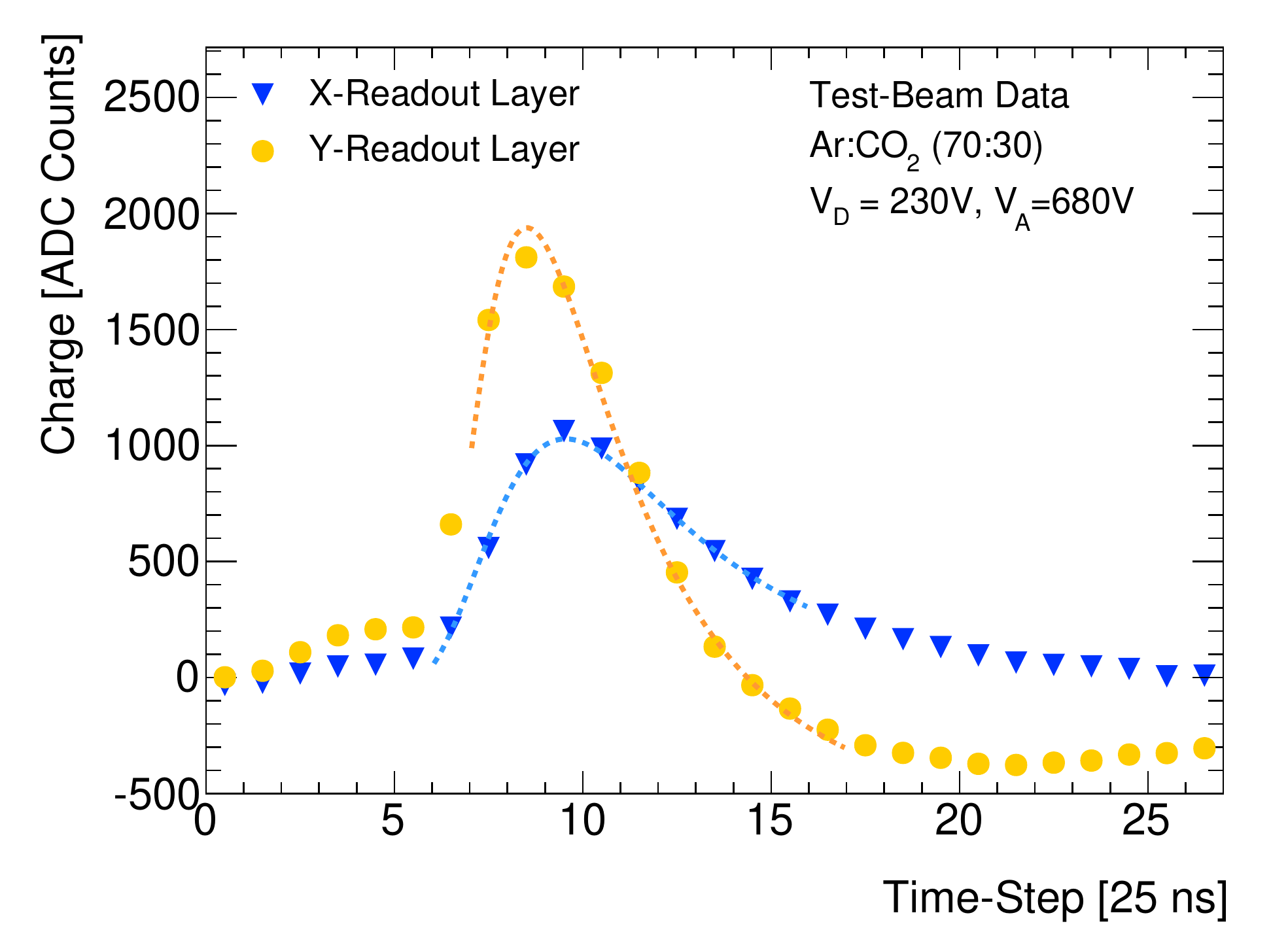}}
\caption{Test-beam data: Recorded signal distribution of a perpendicular incident electron for the x-readout layer (top,left) and the y-readout layer (top,right). The horizontal and vertical axes indicate the channel number and time-step, respectively. The measured charge signal at a given channel and time is encoded in the color spectrum. The measured signal distributions for the readout-strip containing the maximum recorded signal (bottom,left) and the time evolution of charge with single readout strip per event are also shown (bottom,right).}
\label{fig:EventDisplayTestBeam}
\end{figure*}

In order to study the performance of the detector under many different operation conditions, an X-ray generator (Amptek Mini-X) was used to generate highly energetic electrons within the drift-region, via the photoelectric effect. A random trigger has been used during the data-taking. The chamber performance has been studied for different 
amplification voltages, drift voltages and gas-mixtures. We focus our discussion on the dependence on the amplification voltage. It will be shown in the following section that the basic feature of the signal characteristics in comparison between the two readout-layers, do not depend on the gas-mixture. We therefore present our findings only for the X-Ray measurements based on a 93:07 Ar:$\mathrm{CO_2}$ gas mixture and the test-beam measurements based on an Ar:$\mathrm{CO_2}$ gas mixture of 70:30.

The measured distribution of reconstructed events during the X-ray runs over the readout layers in both directions is shown in Figure \ref{fig:EventDisplayPillar}. The presence of the support pillars of the amplification mesh is clearly visible, as they reduce significantly the efficieny for events in their vicinity.

\section{\label{Sec:Signal}Signal Characteristics}

A typical event of the test-beam measurement is shown in Figure \ref{fig:EventDisplayTestBeam}. Because of the CR-RC pulse shaping network in the APV25 circuit, in the case of high count rate, the baseline shift with CR-RC undershoot become non-negligible \cite{APVCMS}. Therefore a negative charge distribution is observed.  The measured signal in neighboring readout strips for 27 time-steps of $25\,$ns lengths are shown for both readout layers. In addition, the measured signal evolution vs. time is shown for the readout strip that contains the maximal recorded signal and also the signal distribution over readout strips for the time step containing the maximal recorded signal is shown for both layers. While the measured signal in a given time-step can be described by a Gaussian function, the time-evolution can be phenomenologically described in the bulk part by a Landau function as shown by the fits in the two lower plots of figure~\ref{fig:EventDisplayTestBeam}.

\begin{figure}[h]
\resizebox{0.495\textwidth}{!}{\includegraphics{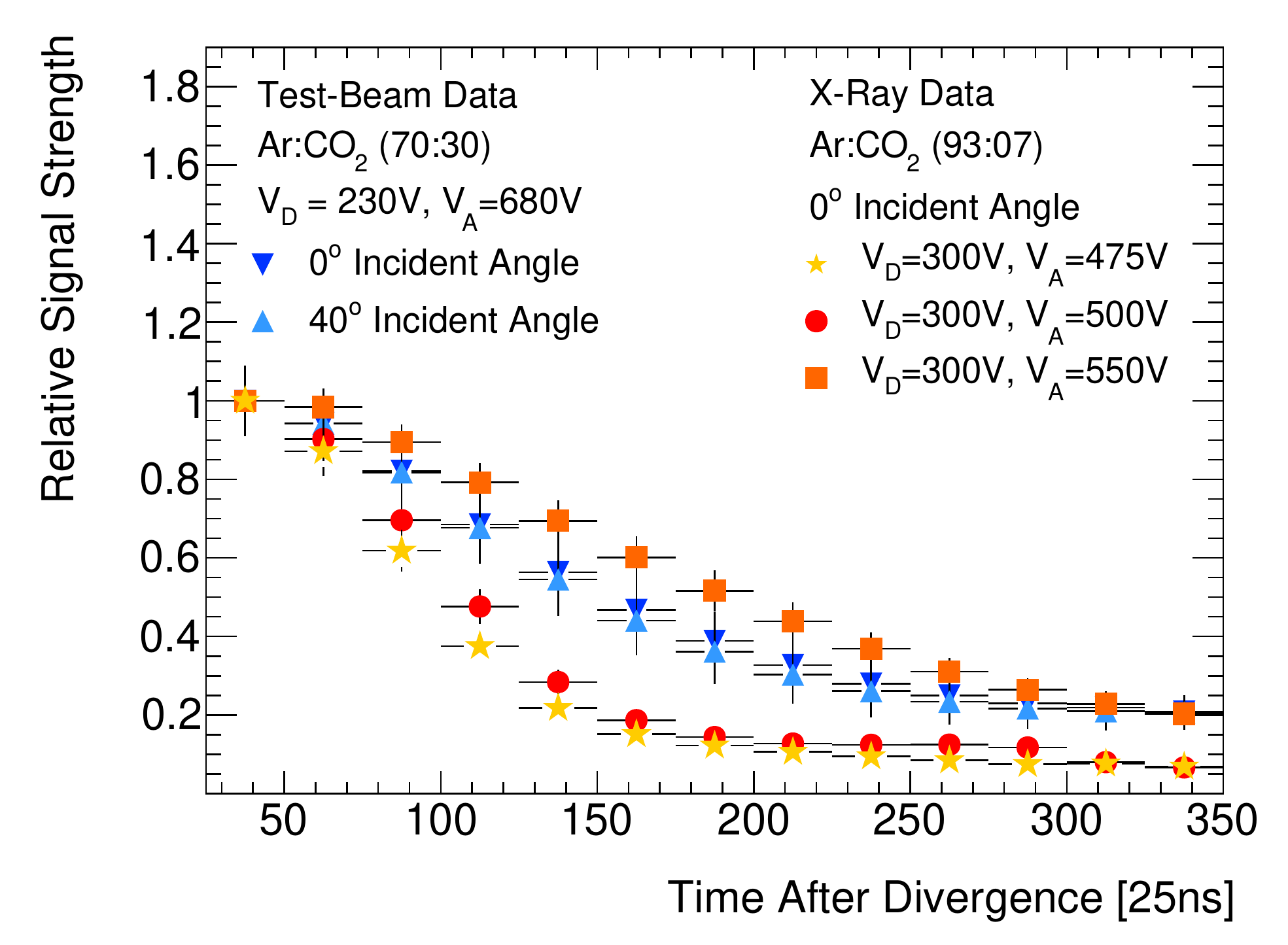}}
\resizebox{0.495\textwidth}{!}{\includegraphics{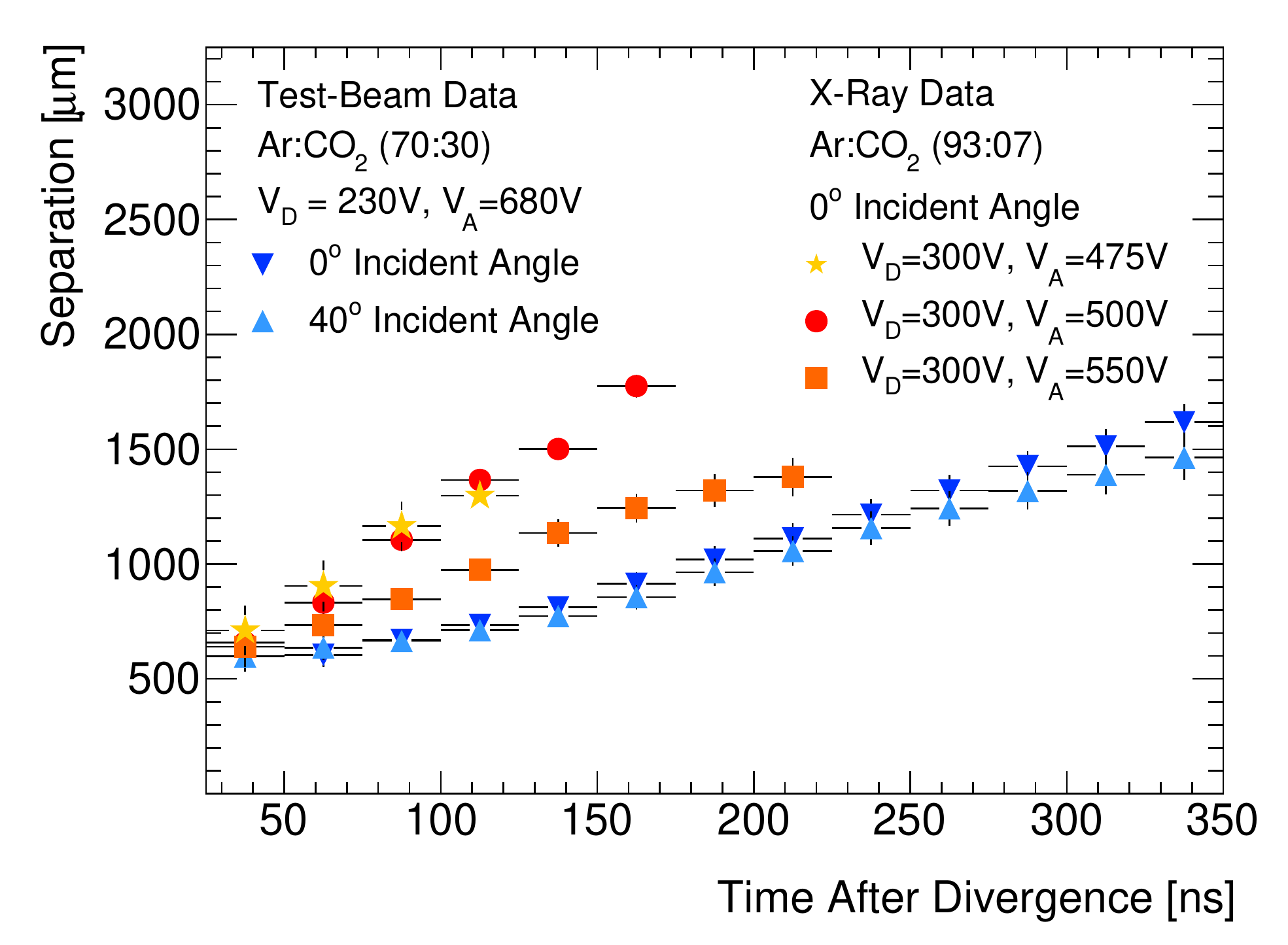}}
\caption{Comparison of the V-shape for two Testbeam measurements with different incident angles and the nominal X-Ray measurement: The signal-strength of the V-shape vs. time (top) and the separation in $\mu m$ of the two signal-peaks vs. time (bottom)}
\label{fig:VShape}
\end{figure}

The measured signal is significantly larger in the y direction, compared to the x direction. This is due to the larger distance of the x layer to the resistive strips and the resulting weaker capacitive coupling. The larger strip width cannot fully accommodate for this effect. In addition, the measured cluster size is localized in three to four strips for the x layer, while it affects more than seven strips of the y-layer. This is closely related with the signal duration, which is significantly longer for the x layer, where the full signal evolution spans over more than $300\,\mathrm{ns}$, while the signal in the y layer lasts not longer than $200\,\mathrm{ns}$. Both differences can be explained by the 'V'-like shape that is seen in the y layer (Figure \ref{fig:EventDisplayTestBeam}). It should be noted that also signal on the y layer is propagated along the resistive strips.

\begin{figure}[h]
\resizebox{0.495\textwidth}{!}{\includegraphics{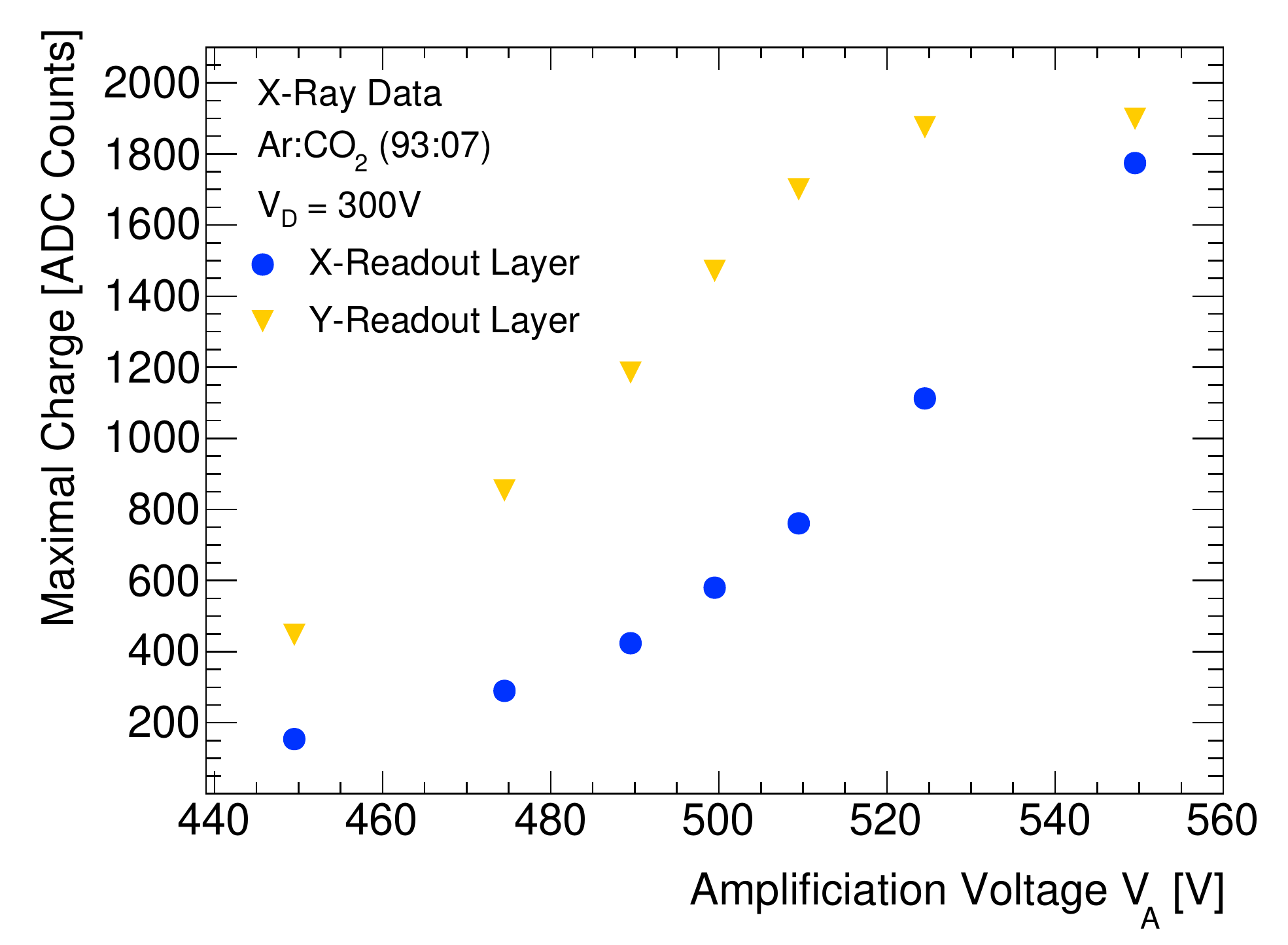}}
\resizebox{0.495\textwidth}{!}{\includegraphics{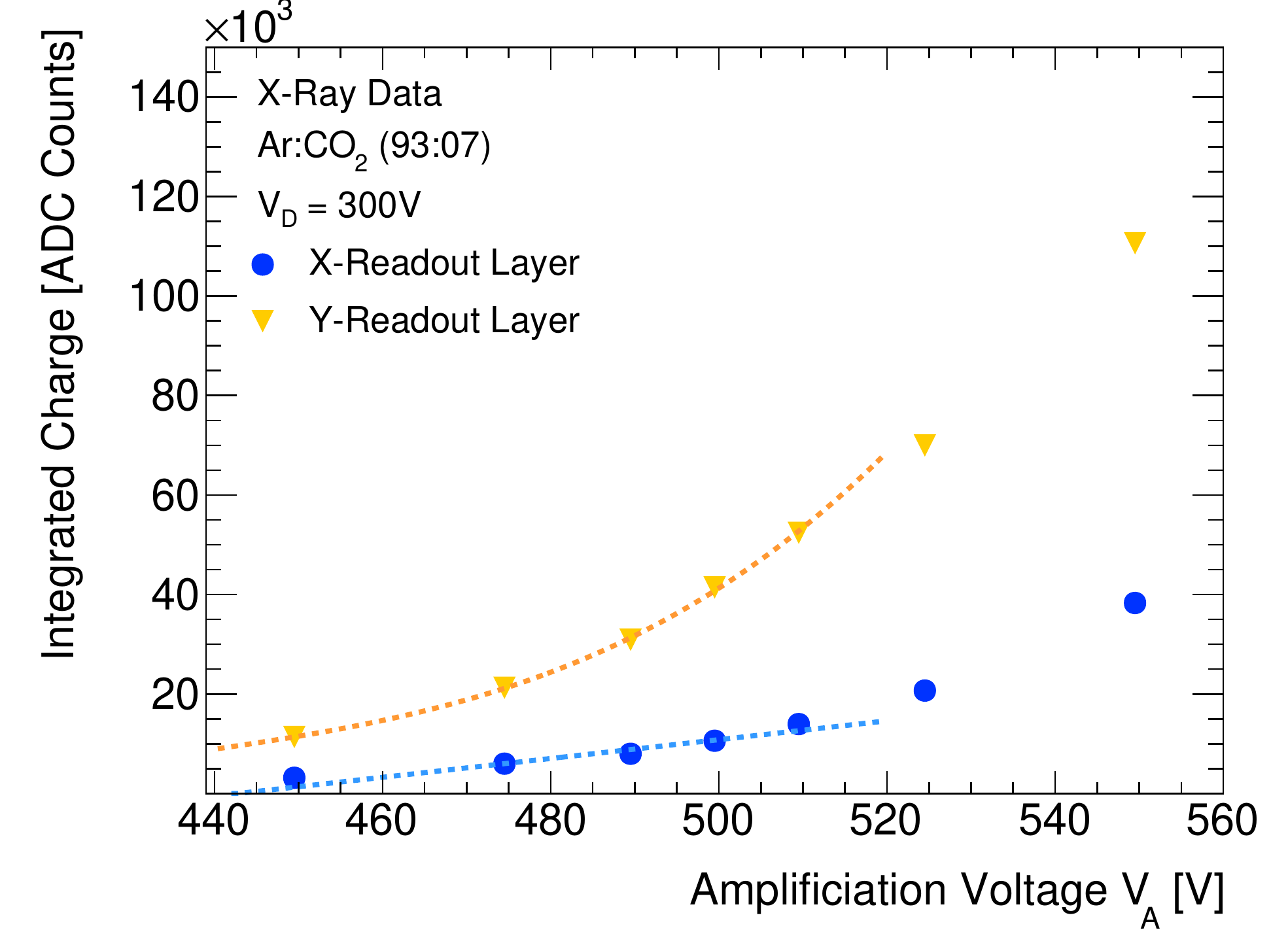}}
\caption{Dependence of signal shape parameters on the amplification voltage based on X-Ray Measurements: signal height (top) and integrated signal (bottom).}
\label{fig:SignalShape1}
\end{figure}

\begin{figure}[h]
\resizebox{0.495\textwidth}{!}{\includegraphics{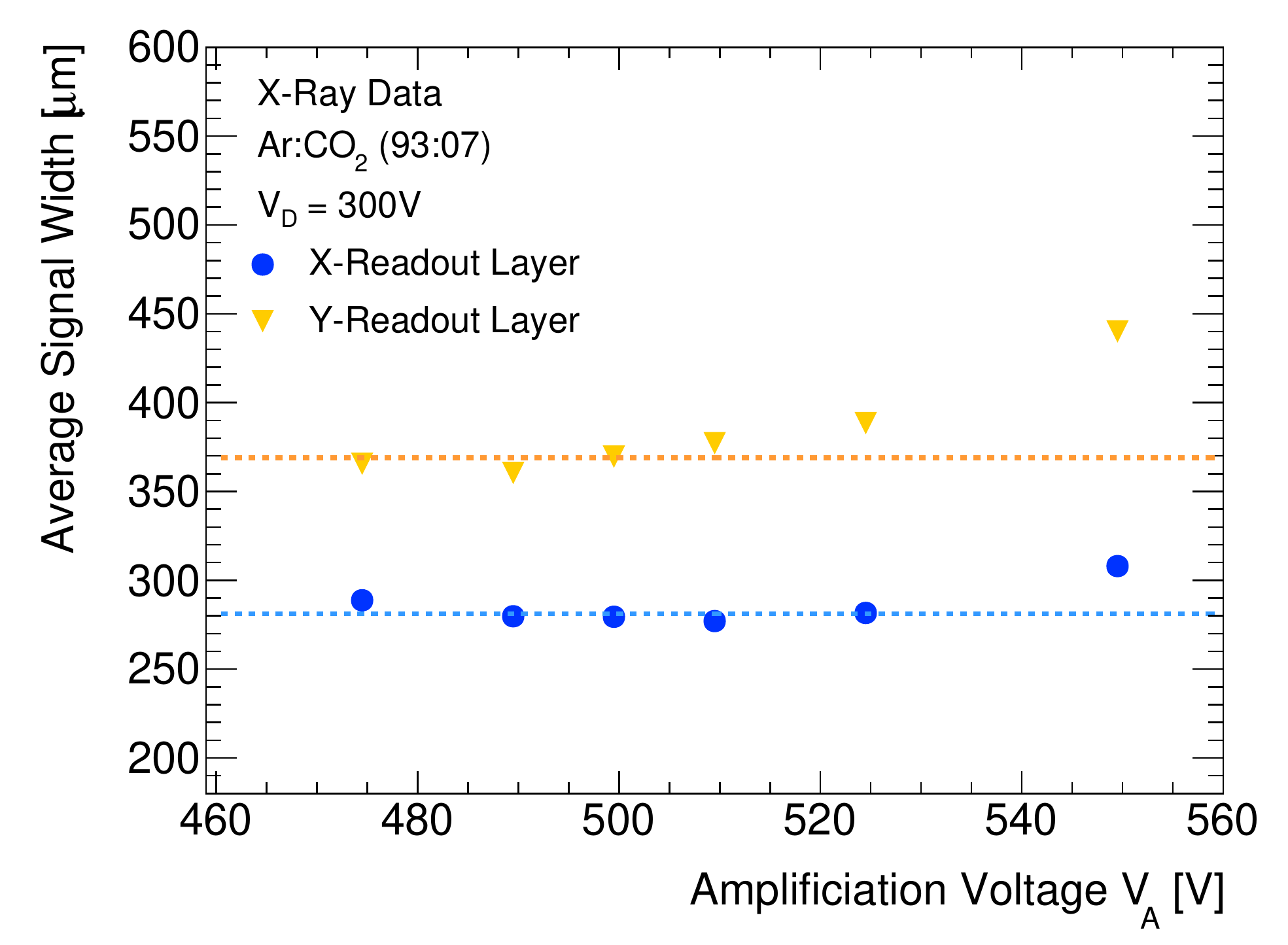}}
\resizebox{0.495\textwidth}{!}{\includegraphics{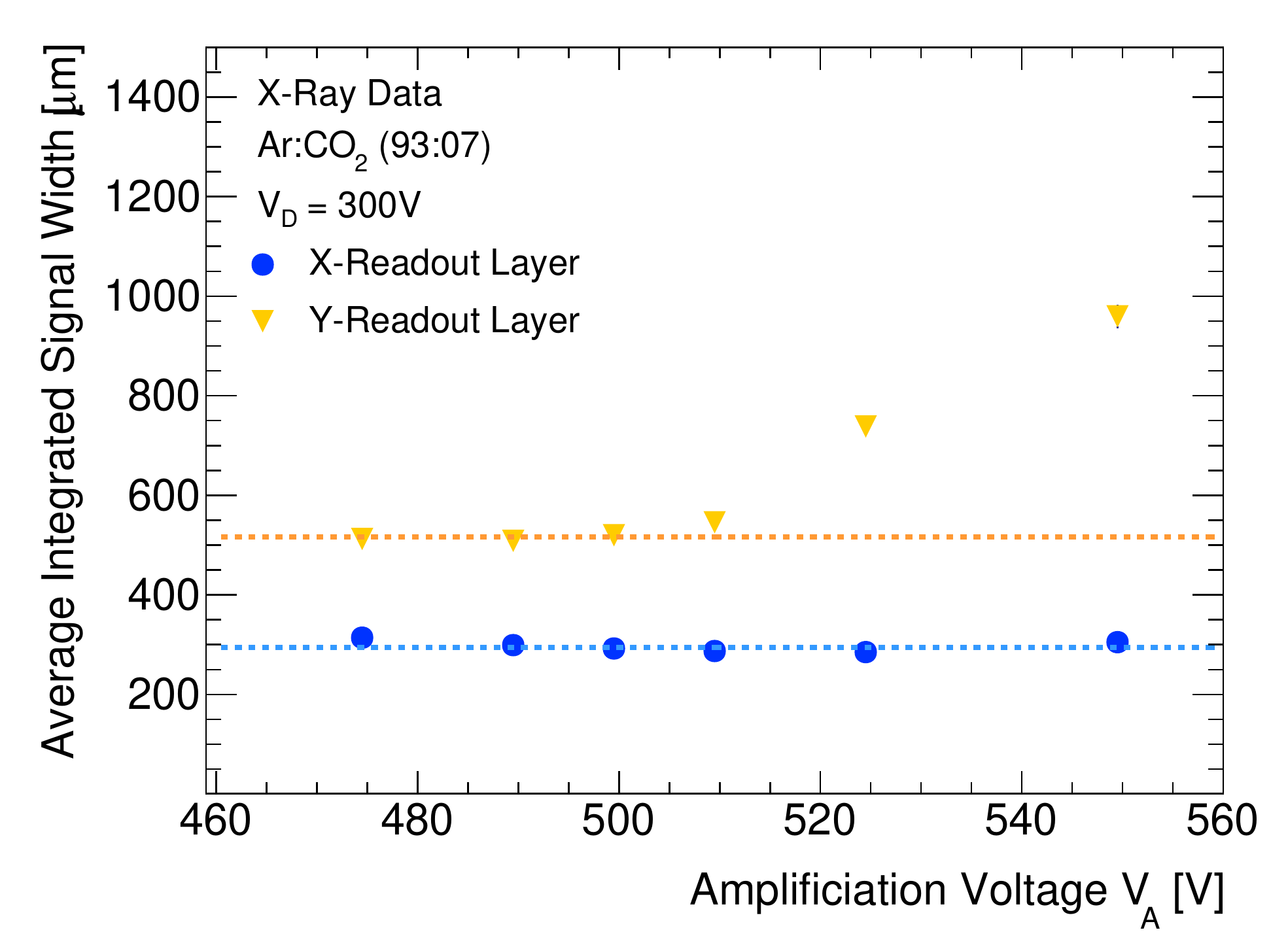}}
\caption{Dependence of signal shape parameters on the amplification voltage based on X-Ray Measurements: signal width at the time-step with the maximum signal (top) and signal width integrated over time (bottom)}
\label{fig:SignalShape2}
\end{figure}

\begin{figure}[h]
\resizebox{0.495\textwidth}{!}{\includegraphics{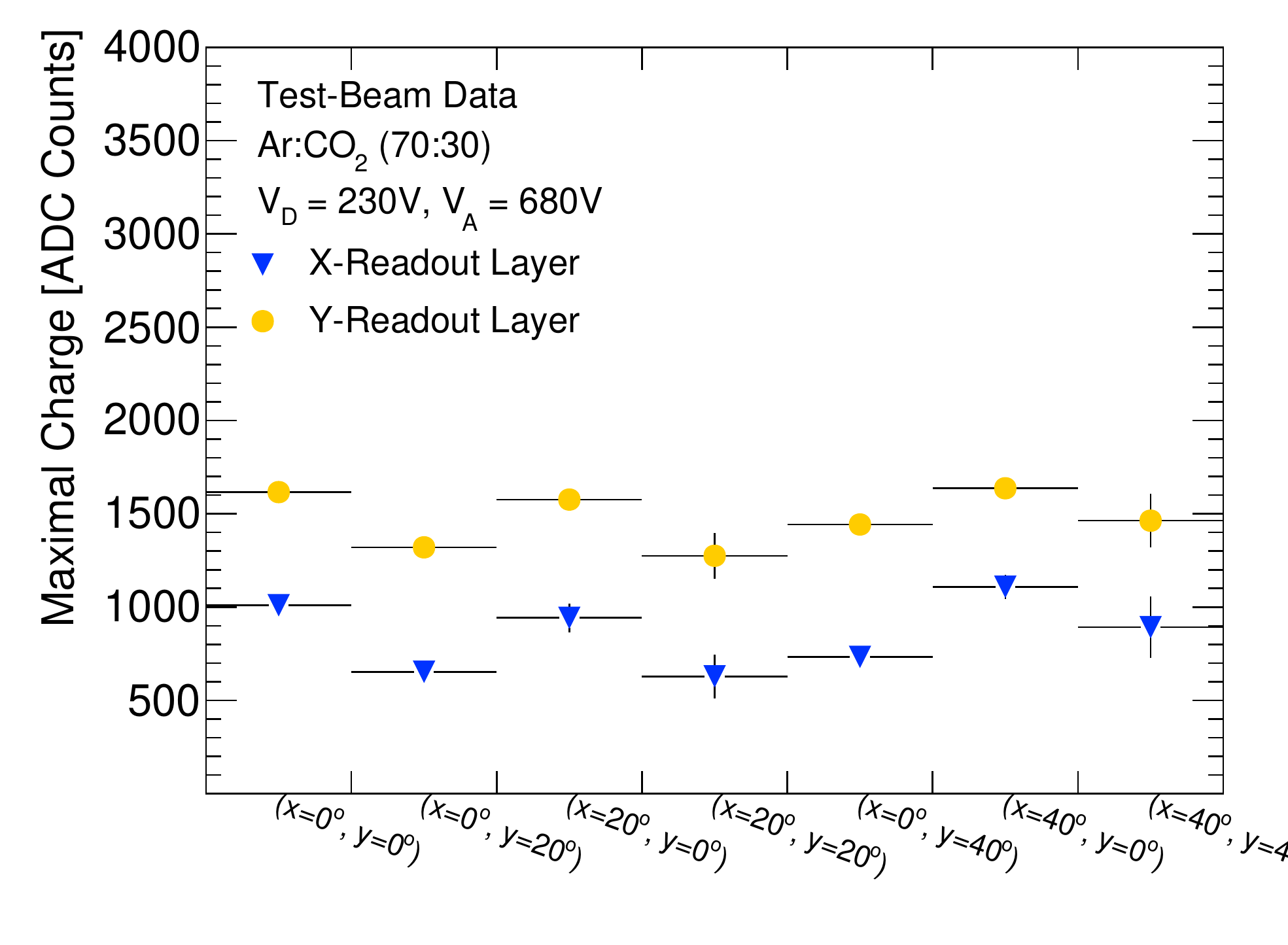}}
\resizebox{0.495\textwidth}{!}{\includegraphics{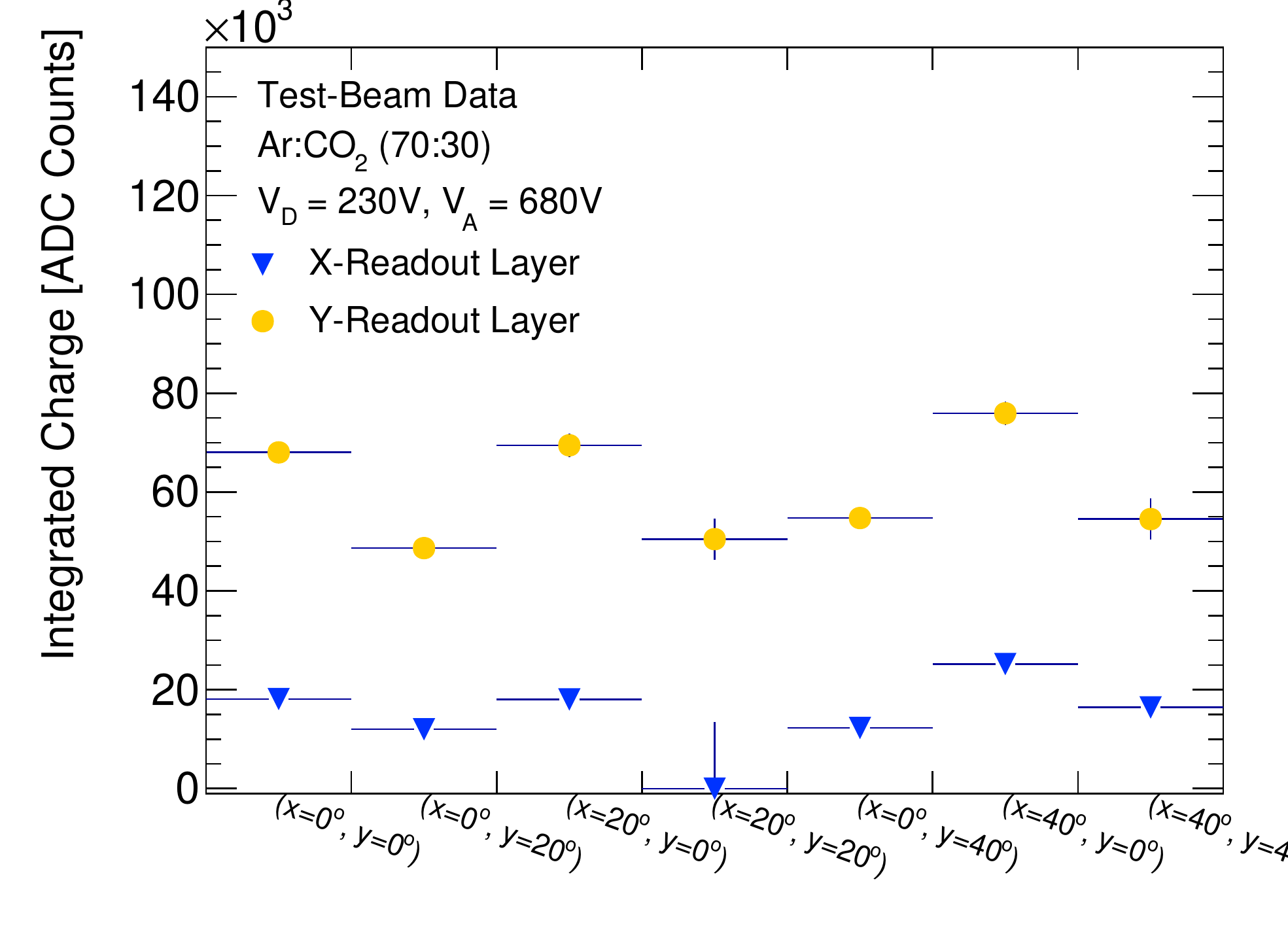}}
\caption{Dependence of signal height (top) and the integrated charge (and) on the incident angle of incoming particles based on test-beam data.}
\label{fig:SignalShape3}
\end{figure}


Before discussing the V~shape in more detail, the typical event shapes of the X-Ray measurement series should be discussed. 
Similar features that have been present in the test-beam data can also been seen here, i.e. a larger signal strength in the y-layer that also affects more readout-strips and evolves longer in time. In addition, also the V~shape of the signal is observed in each event.

The initial signal of the V~shape signature in the y-layer is measured on 3-4 neighboring strips within the first $100\,\ns$. This is comparable to the signal of the x layer. However, the incident charge starts to spread along the resistive strips, i.e. perpendicular to the y layer strips and hence creates a tail via capacitive coupling. Similar effect are therefore expected to be present in all Micromegas chambers where the resistive strips are not placed parallel to the readout-strips. 

It is worth while to study the V~shape signature in more detail, since it affects more channels and leads to longer signal times, which in turn reduces the capability of chambers for operation at high rate. Figure \ref{fig:VShape} shows the relative signal height of the V shape vs. time for several X-Ray measurements and two test-beam measurements for different incident angles, where the signal was normalized to the maximum charge. In addition, the distance of the two signal peaks of the V~shape vs. time is shown. Both parameters show no dependence on the incident angle. Furthermore, the distance between the signal peaks increases linearly with time for all measurements. It should be noted that the speed of signal height separation in the V~shape depends on the amplification voltage, i.e. the higher $V_A$, the lower the separation. This effect is therefore correlated with the larger gain and hence to the larger signal which is induced at the resistive strips.

Having described the basic signal properties in both readout-layers, their dependencies on variable operation parameters like amplification voltage, drift-voltage, gas-mixture and incident angle of incoming particles can be discussed. The signal properties under study have been the signal height, i.e. the maximum recorded charge, the spatial signal width (defined as a Gaussian width) during the time-step with the highest recorded charge, and the integrated signal over all affected channels during the full time-evolution. While the dependencies on different drift and amplification voltages have been studies with the X-Ray measurements, the dependency studies on different incident angles are based on the test-beam dataset. 

A summary of our results for the amplification voltage dependencies can be seen in Figures \ref{fig:SignalShape1} and \ref{fig:SignalShape2}.  The signal height is expected to increase linearly with increasing amplification voltage $V_A$ up to a certain $V_A$. For $V_A\geq 510\,\mathrm{V}$, the signal height of the y-layer saturates, which is due to the limitations of our readout-system. The saturation point of the lower x-layer has not been reached within our studies. However, below $V_A<500$ the increase in signal height is lower for the x-readout layer. This contradicts a naive picture, where the increase in charge in the resistive strips should have an equal effect on both readout-strips. One possible explanation could be charge screening effects of the x-readout layer w.r.t. the lower y-layer. Also a different dependence on $V_A$ can be seen for the integrated charge for both layers, which do not longer exhibit a linear dependence on $V_A$. The width of the signal distribution at the time-step which contains the maximum charge is fully independent of $V_A$ before the saturation. This behaviour is also expected as the higher voltage in the amplification volume does not impact the electron/ion spread which is mainly caused in the drift-region. The increase of the width for $V_A>510\,V$ is an artefact of the Gaussian fit of the signal, since the signal height gets constant over neighboring strips when reaching the readout saturation. A similar behaviour can be seen for the width of the integrated charge distribution over the full time. 

A significant dependence of the signal shape on the incident angle has been observed in the test-beam data (Figure \ref{fig:SignalShape3}). Different incident angles w.r.t. to the y-layer lead to a significant reduction of the signal height in both layers, while we observe no reduction for incident angles w.r.t. the x-layer. However, it should be noted that the ratio of signal heights between both layers is constant.

In summary, we see a different response of both layers w.r.t. to variations of the amplification voltage and the incident angle of particles. These differences can be explained by the differences in the capacitive coupling between the resistive strips that are parallel for one readout-layer but perpendicular for the second. 

\section{\label{Sec:Efficiency}Efficiency Measurements}

In the previous section, the differences in the response of the two readout layers have been discussed. The main question is clearly wether these differences in the signal characteristics lead to a change of the subsequent performance parameters, such as the signal reconstruction efficiency of the chamber. In order to study this aspect, we define a signal as 'loosely' reconstructable, if it shows a pronounced peak within a $75\,\mathrm{ns}$ time-window around the expected peak arrival time and a peak clearly above noise level. A 'tightly' reconstructable signal is defined as a 'loose' signal that has at least two neighboring readout strips in both directions with decreasing signal strength and also a decreasing signal strength recorded two time-steps of $25\,\mathrm{ns}$ before and after the maximal recorded charge. This decreasing behavior in spatial- and time-direction is important to achieve a precise position- and time-measurement of the recorded event.

While 'loosely' reconstructed events can consist of a single signal in one readout strip, a 'tightly' reconstructed peak has a peak shape form in spatial and time direction. It should be noted that signals in the saturation regime of the electronics will not lead to a 'tightly' reconstructed event, as neighbouring cells can have the same height and are thefore not Gaussian but have a plateau type form.

\begin{figure}[ht]
\resizebox{0.495\textwidth}{!}{\includegraphics{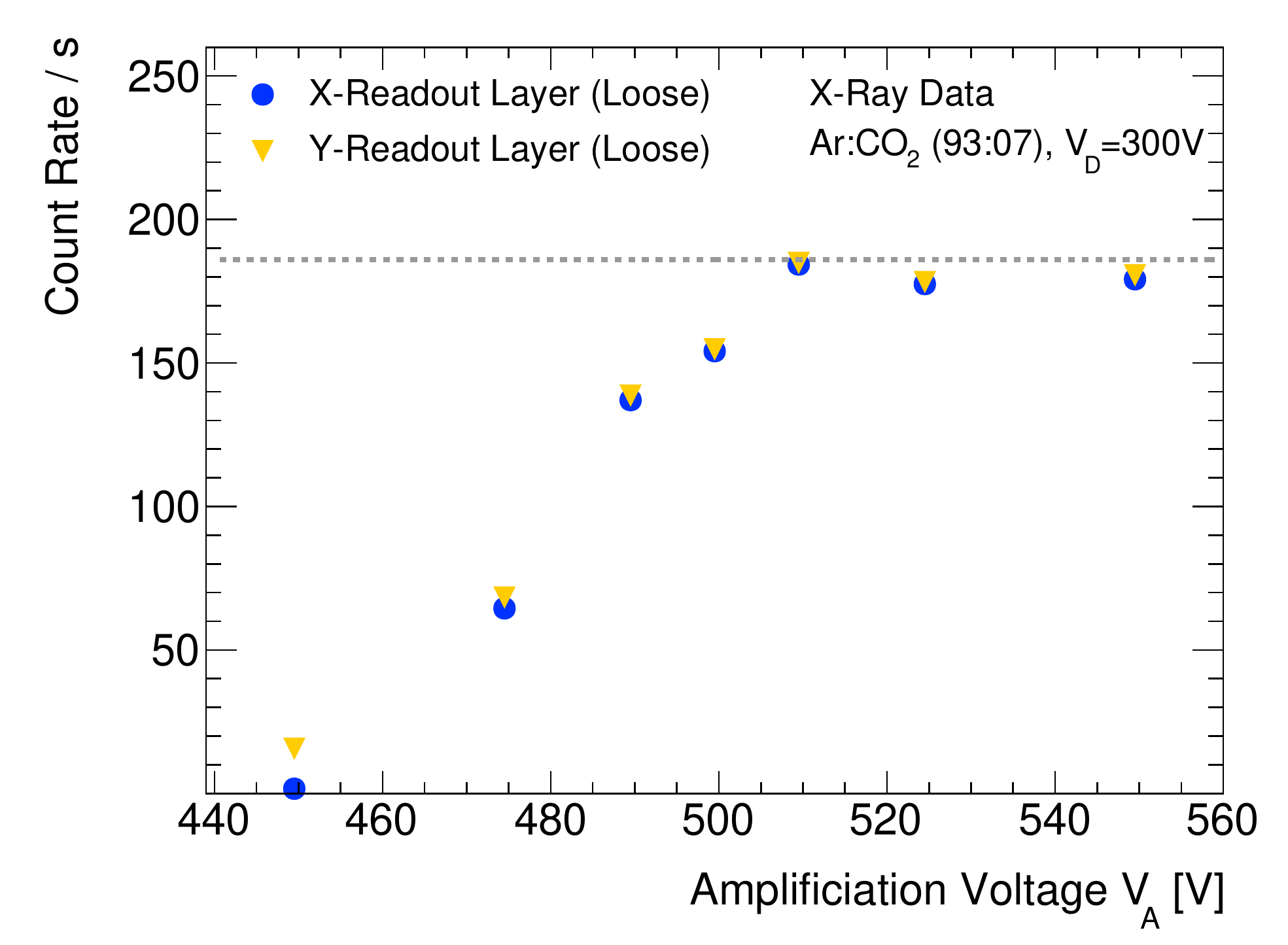}}
\resizebox{0.495\textwidth}{!}{\includegraphics{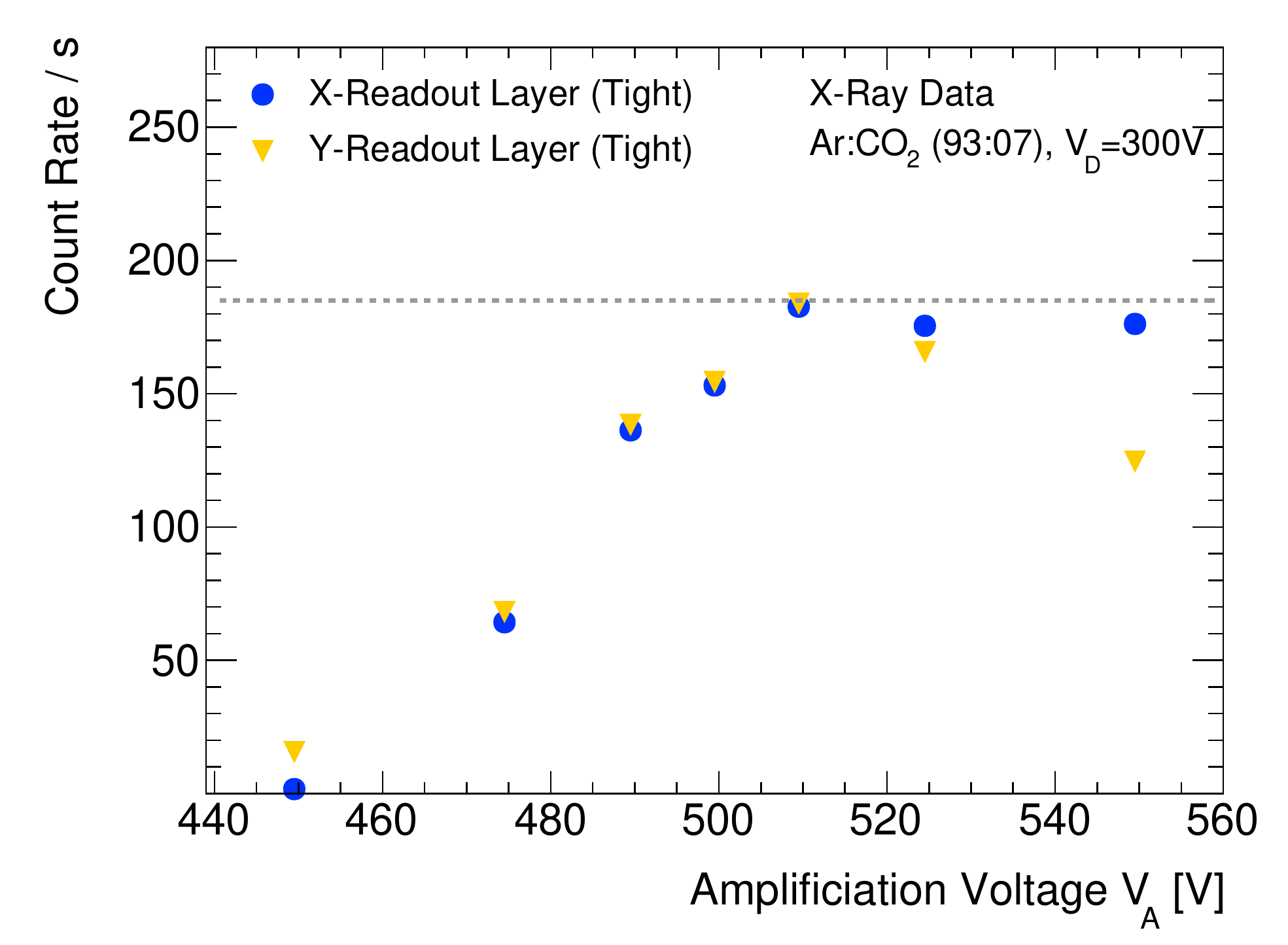}}
\caption{The counting rate of loosely (top) and tightly (bottom) reconstructed X-ray events per second for different amplification voltages $V_A$.}
\label{fig:CountRate}
\end{figure}

In a first step, we test the counting rate of 'loosely' and 'tightly' reconstructed events independently in the two readout layers for different amplification voltages. The results are illustrated in Figure \ref{fig:CountRate}. Both layers show a very similar behavior. In particular, we observe a strong increase of the counting rate with increasing $V_A$ up to the saturation regime. At too low amplification voltage (eg. $475\, \mathrm{V}$), the counting rate in x~layer is slightly lower than in y~layer due to the x-strips get larger charge coupling length. The results are different for the 'tight' reconstruction criteria. While the upper y~layer shows in general a higher counting rate compared to the x~layer, we also observe a decrease for $V_A$ in the saturation regime. This can be explained by the plateau type form of the measured signal for saturated events.

In a second step, we study the efficiency of one layer with respect to the second layer, by first searching for a pronounced peak in one layer. This peak is then used to define the time of the maximum recorded signal by fitting a Landau function to the measured distribution. In a second step, a pronounced peak in the second layer is searched, that appears within a time-window of $75\,\mathrm{ns}$. It turns out, that the maximal charge is reached at the same time for both layers. The width of the difference of the measured peak-times of both layers is $15\,\mathrm{ns}$, implying a timing resolution of $15\,\ns/\sqrt{2}$ in the most optimal case per layer.

\begin{figure}[t]
\resizebox{0.495\textwidth}{!}{\includegraphics{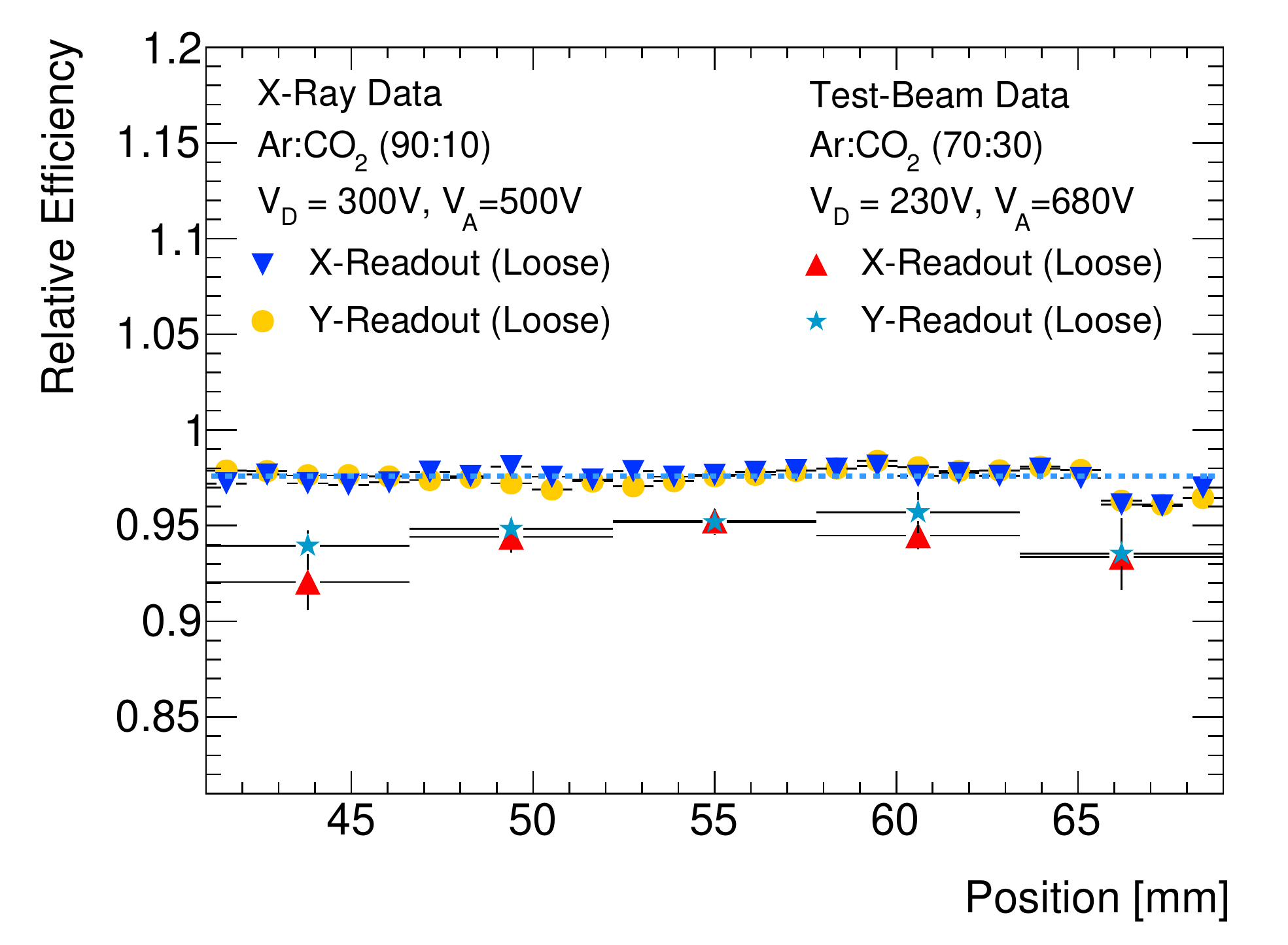}}
\resizebox{0.495\textwidth}{!}{\includegraphics{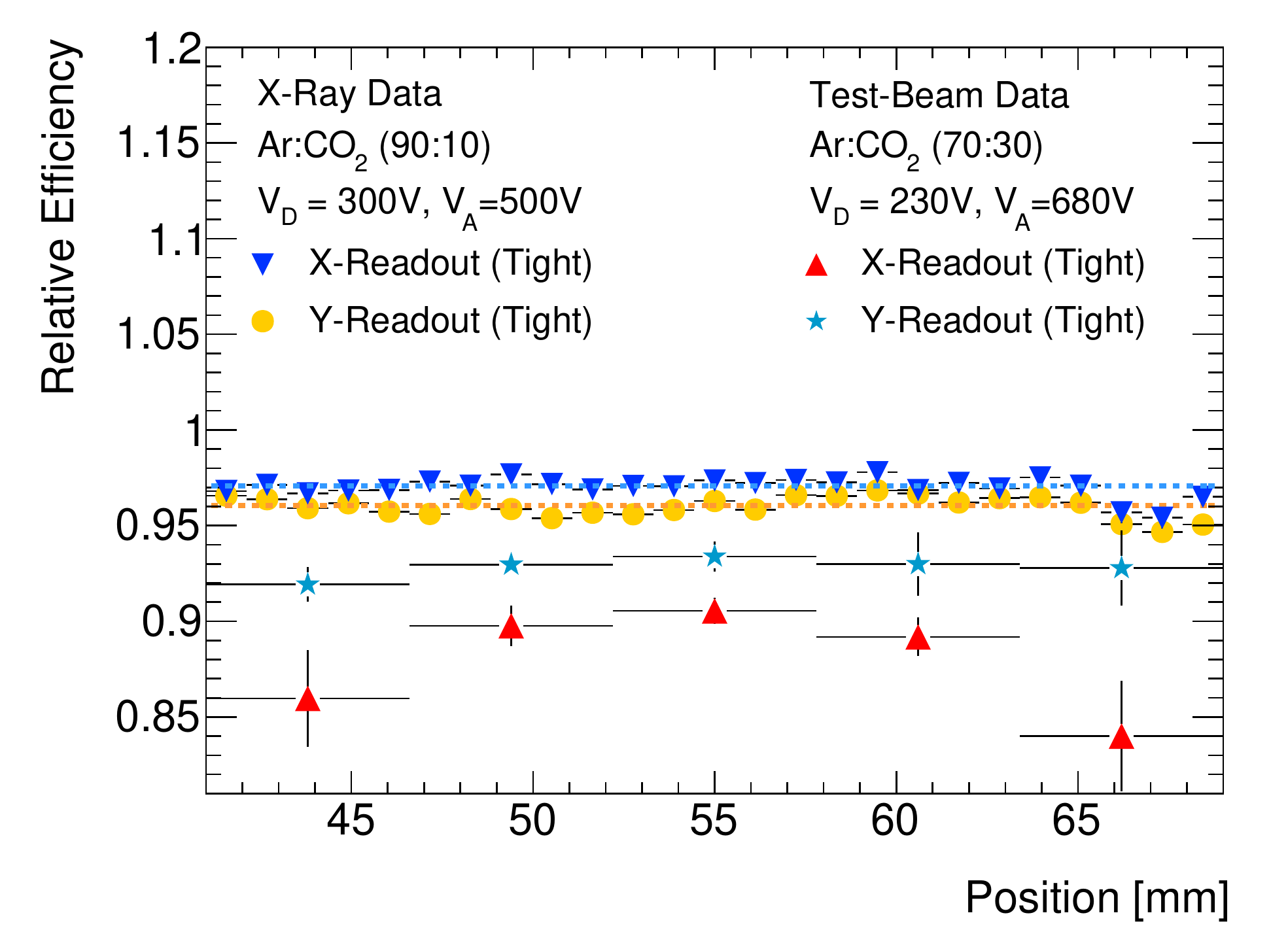}}
\caption{Relative reconstruction efficiency of both layers in dependence of the position on the readout strip for loose (top) and tight (bottom) conditions for test-beam and X-ray data.}
\label{fig:EffCompXLayer1}
\end{figure}

\begin{figure}[t]
\resizebox{0.495\textwidth}{!}{\includegraphics{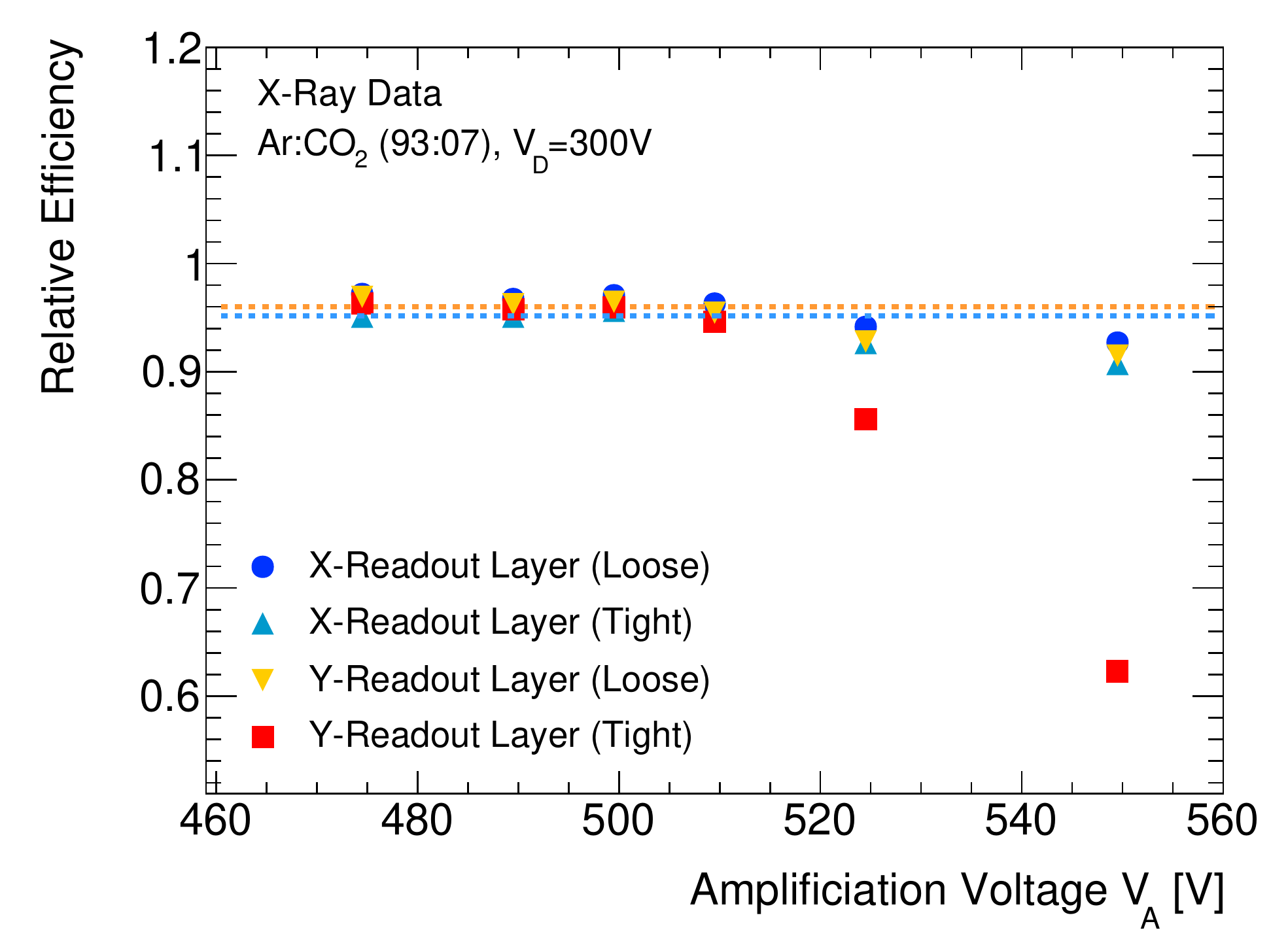}}
\resizebox{0.495\textwidth}{!}{\includegraphics{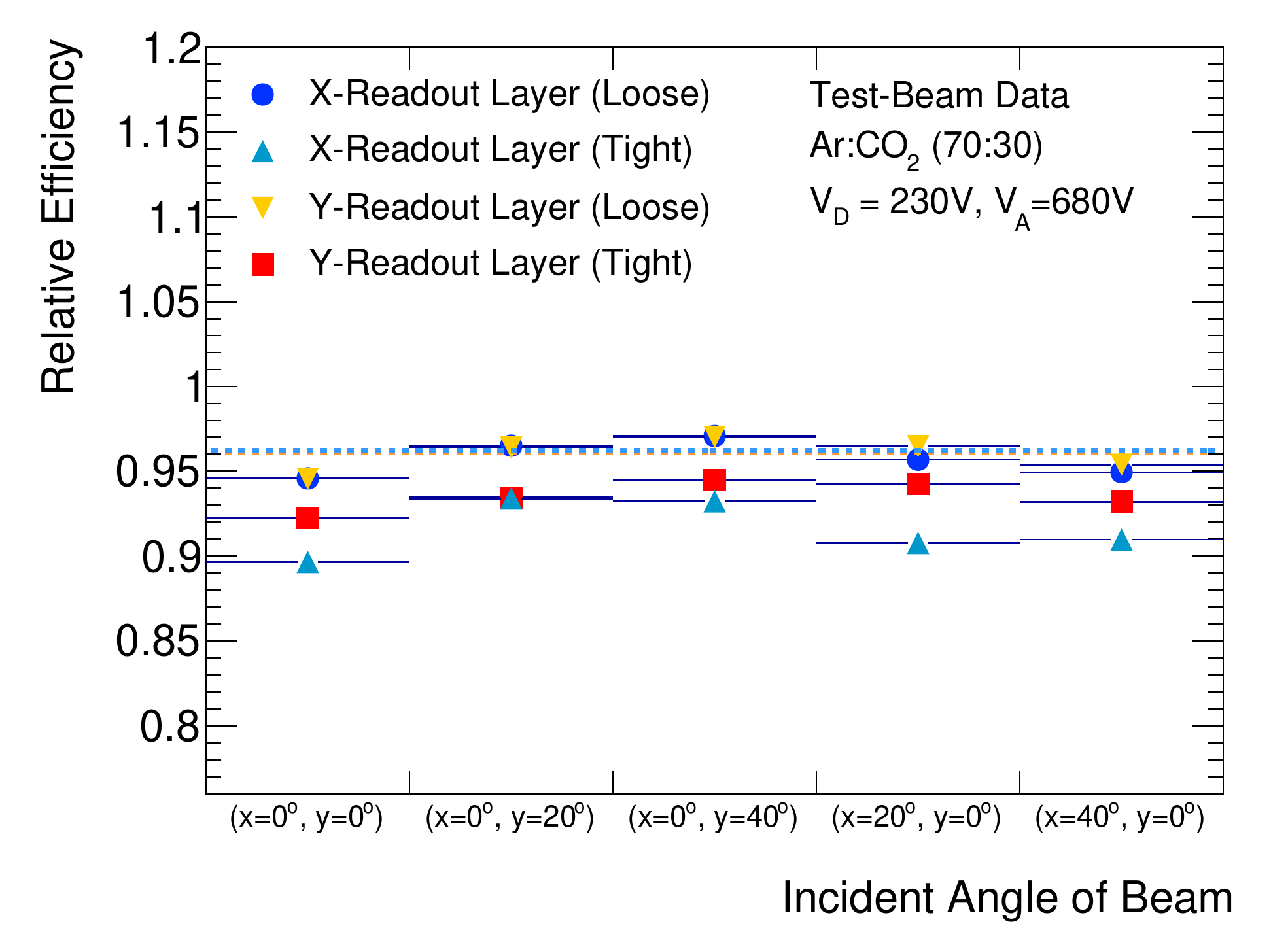}}
\caption{Relative reconstruction efficiency of both layers in dependence of the amplification voltage (top) and the incident angle (bottom).}
\label{fig:EffCompXLayer}
\end{figure}

The resulting relative efficiencies for different positions on the readout-strip are shown in Figure \ref{fig:EffCompXLayer1} for 'loose' and 'tight' reconstruction conditions. First of all, it should be noted, that the average efficiency of the test-beam data is significantly lower than for the X-ray results. This is due to the sub-optimal operation conditions of the Micromegas chamber during the test-beam run, i.e.~due to the sub-optimal gas mixture and the high amplification voltage, leading to charge saturation and a lower reconstruction efficiency. In general, the reconstruction efficiency of the y~layer is slightly higher than the x~layer, but both efficiencies turn out to be independentof the strip position. This is an important result, as it highlights the independence of relative efficiency from the support pillar structure. The dependence of the relative efficiencies on the amplification voltage and the incident beam angle are displayed in Figure \ref{fig:EffCompXLayer}. We see a small dependence on the incident beam angle and also no dependence on the amplification voltage below $V_A<510\,V$. This is an important result, as it underlines the stability of the Micromegas detector with a two-dimensional readout structure across varying operation and incident particle conditions. The drop of 'tight' reconstruction efficiency on the x-readout layer for higher amplification voltages can again be explained by the signal shape in the saturation regime.

\section{\label{Sec:Summary}Summary and Conclusion}

In this paper, we have summarized the signal characteristics of a 2D Micromegas detector in dependence of the applied amplification and drift voltages and the gas mixture used, based on test-beam and X-ray data. 

While it was suggested in previous studies that a properly adjusted width of the readout strips can equalise the charge response of both layers, our measurements show that the effect of a change in the amplification voltage on the charge response is different for the two layers. In addition, we show that the choice of two perpendicular readout layers leads automatically to a different signal response in both layers. However, no significant effect on the detector performance itself could be seen. In future test-beam measurements, it is foreseen to further study the effect of different operation conditions on the spatial resolution. Only small effects are expected here, since we already show that the signal width is independent of the operation conditions.

\section*{Acknowledgements}

We would like to acknowledge the close collaboration with Rui de Oliveira from the CERN PCB workshop where the detector prototype was built. In addition, we would like to thank Alexopoulos Theodoros and Frank Fiedler for the useful comments during the preparation of this paper, as well as the MAMI accelerator team, Matthias Hoeck and Pepe Guelker for their help during our test-beam measurements. This work is supported by the Volkswagen Foundation and the German Research Foundation (DFG). 

\bibliography{Micromegas2DPerformance}

\begin{thebibliography}{10}
\expandafter\ifx\csname url\endcsname\relax
  \def\url#1{\texttt{#1}}\fi
\expandafter\ifx\csname urlprefix\endcsname\relax\def\urlprefix{URL }\fi
\expandafter\ifx\csname href\endcsname\relax
  \def\href#1#2{#2} \def\path#1{#1}\fi

\bibitem{Giomataris:1995fq}
Y.~Giomataris, P.~Rebourgeard, J.~Robert, G.~Charpak, {MICROMEGAS: A High
  granularity position sensitive gaseous detector for high particle flux
  environments}, Nucl.Instrum.Meth. A376 (1996) 29--35.
\newblock \href {http://dx.doi.org/10.1016/0168-9002(96)00175-1}
  {\path{doi:10.1016/0168-9002(96)00175-1}}.

\bibitem{Giomataris:2004aa}
I.~Giomataris, R.~De~Oliveira, S.~Andriamonje, S.~Aune, G.~Charpak, et~al.,
  {Micromegas in a bulk}, Nucl.Instrum.Meth. A560 (2006) 405--408.
\newblock \href {http://arxiv.org/abs/physics/0501003}
  {\path{arXiv:physics/0501003}}, \href
  {http://dx.doi.org/10.1016/j.nima.2005.12.222}
  {\path{doi:10.1016/j.nima.2005.12.222}}.

\bibitem{Alexopoulos:2009zza}
T.~Alexopoulos, A.~Altintas, M.~Alviggi, M.~Arik, S.~Cetin, et~al., {The ATLAS
  muon Micromegas \& project: Towards large-size chambers for the s-LHC}, JINST
  4 (2009) P12015.
\newblock \href {http://dx.doi.org/10.1088/1748-0221/4/12/P12015}
  {\path{doi:10.1088/1748-0221/4/12/P12015}}.

\bibitem{Wotschack:2013ola}
J.~Wotschack, {The development of large-area Micromegas detectors for the ATLAS
  upgrade}, Mod.Phys.Lett. A28 (2013) 1340020.
\newblock \href {http://dx.doi.org/10.1142/S0217732313400208}
  {\path{doi:10.1142/S0217732313400208}}.

\bibitem{Chefdeville:2011zz}
M.~Chefdeville, {RD51, a world-wide collaboration for the development of micro
  pattern gaseous detectors}, J.Phys.Conf.Ser. 309 (2011) 012017.
\newblock \href {http://dx.doi.org/10.1088/1742-6596/309/1/012017}
  {\path{doi:10.1088/1742-6596/309/1/012017}}.

\bibitem{Alexopoulos:2010zz}
T.~Alexopoulos, A.~Altintas, M.~Alviggi, M.~Arik, S.~Cetin, et~al.,
  {Development of large size Micromegas detector for the upgrade of the ATLAS
  muon system}, Nucl.Instrum.Meth. A617 (2010) 161--165.
\newblock \href {http://dx.doi.org/10.1016/j.nima.2009.06.113}
  {\path{doi:10.1016/j.nima.2009.06.113}}.

\bibitem{Alexopoulos:2011zz}
T.~Alexopoulos, J.~Burnens, R.~de~Oliveira, G.~Glonti, O.~Pizzirusso, et~al.,
  {A spark-resistant bulk-micromegas chamber for high-rate applications},
  Nucl.Instrum.Meth. A640 (2011) 110--118.
\newblock \href {http://dx.doi.org/10.1016/j.nima.2011.03.025}
  {\path{doi:10.1016/j.nima.2011.03.025}}.

\bibitem{Dris:2014qpa}
M.~Dris, T.~Alexopoulos, {Signal Formation in Various Detectors}\href
  {http://arxiv.org/abs/1406.3217} {\path{arXiv:1406.3217}}.

\bibitem{Byszewski:2012zz}
M.~Byszewski, J.~Wotschack, {Resistive-strips micromegas detectors with
  two-dimensional readout}, JINST 7 (2012) C02060.
\newblock \href {http://dx.doi.org/10.1088/1748-0221/7/02/C02060}
  {\path{doi:10.1088/1748-0221/7/02/C02060}}.

\bibitem{Martoiu:2013aca}
S.~Martoiu, H.~Muller, A.~Tarazona, J.~Toledo, {Development of the scalable
  readout system for micro-pattern gas detectors and other applications}, JINST
  8 (2013) C03015.
\newblock \href {http://dx.doi.org/10.1088/1748-0221/8/03/C03015}
  {\path{doi:10.1088/1748-0221/8/03/C03015}}.

\bibitem{Martoiu:2011zja}
S.~Martoiu, H.~Muller, J.~Toledo, {Front-end electronics for the Scalable
  Readout System of RD51}, IEEE Nucl.Sci.Symp.Conf.Rec. 2011 (2011) 2036--2038.
\newblock \href {http://dx.doi.org/10.1109/NSSMIC.2011.6154414}
  {\path{doi:10.1109/NSSMIC.2011.6154414}}.

\bibitem{APVJones}
L.Jones, et~al., {The APV25 deep sub Micron readout chip for CMS channels
  detectors}, Proceedings of 5th workshop on Chips electronics for LHC
  experiments CERN/LHCC/99-09 (1999) 162--166.
\newblock \href {http://dx.doi.org/10.1088/1748-0221/8/03/C03015}
  {\path{doi:10.1088/1748-0221/8/03/C03015}}.

\bibitem{APVCMS}
M.~J. French, L.~L. Jones, Q.~R. Morrissey, A.~Neviani, R.~Turchetta, et~al.,
  {Design and results from the APV25, a deep sub-micron CMOS front-end chip for
  the CMS tracker}, Nucl. Instrum. Methods Phys. Res., A 466 (2001) 359--365.
\newblock \href {http://dx.doi.org/10.1016/S0168-9002(01)00589-7}
  {\path{doi:10.1016/S0168-9002(01)00589-7}}.

\end{thebibliography}

\end{document}